\documentclass[10pt,a4paper]{article}
\usepackage{jheppub}

\begin{document}
\title{Charmless $B_{u,d,s}\to VT$ decays in perturbative QCD approach}
\author[a]{S. Freddy,}
\author[a,1]{C.S. Kim\note{Corresponding author},}
\author[a]{R.H. Li,}
\author[b]{and Z.T. Zou}

\affiliation[a]{Department of Physics $\&$ IPAP, Yonsei University, Seoul 120-479, Korea}
\affiliation[b]{Institute of High Energy Physics, P.O. Box 918(4), Beijing 100049, China}

\emailAdd{cskim@yonsei.ac.kr}

\abstract{
Motivated by the experimental data, we study charmless $B_{u,d,s}\to
VT$ ($V$ and $T$ denote light vector and tensor mesons respectively)
decays in the perturbative QCD approach. The predictions
of branching ratios, polarization fractions and direct CP violations
are given in detail. Specifically, within this approach we have calculated the polarization
fractions and the branching ratios of $B\to \phi(K_2^{*-}, \bar K_2^{*0})$ which
agree well with the observed experimental data, however
the branching ratios of $B\to \omega(K_2^{*-}, \bar K_2^{*0})$ are hard
to be explained, where the polarization fractions are well accommodated.
The tree dominated channels with a vector meson emitted have longitudinal polarization fraction of $90\%$, while the penguin dominating ones have subtle
polarization fractions. Fortunately, most branching ratios of
$B_{u,d}$ decays are of the order $10^{-6}$, which would be straight forward for
experimental observations. For the $B_s$ decays the branching ratios can reach the order of $10^{-6}$ in tree dominated decays,
while in penguin dominated decays those are of order of $10^{-7}$ which
require more experimental data to be observed.
}

\maketitle

\section{Introduction}
Flavor physics has been being thoroughly investigated for many years with the advent of $B$-factories. As more and more
experimental data is accumulated, flavor physics plays an important
role in the precision test of the standard model (SM) and beyond the SM as well as studying the
properties of many light hadrons.  A few $B\to VT$ decay channels have been already
reported by the BaBar collaboration \cite{Aubert:2009sx,Aubert:2008bc,Aubert:2008zza},
which makes the $B$ to tensor meson\footnote{Tensor mesons with $J^P=2^+$ have recently become
one of many hot topics.} decays gain
more and more attention.

Even before the experimental reports there had been already a couple of
works \cite{LopezCastro:1997im,Munoz:1998sn}, which studied the $B \to VT$ decays involving a charmed tensor meson under the quark model. Here we would like to
consider the charmless $B\to VT$ decays instead.  As early works, these charmless
decays had been also studied in the framework of generalized factorization
\cite{Kim:2001py} and in Isgur-Scora-Grinstein-Wise updated
model \cite{Kim:BtoVT}. Later these decays were again
studied in the covariant light-front approach in Ref.
\cite{JHMunoz:2009aba}. Polarizations of $B\to VT$ decays as well are
studied in Ref. \cite{Datta:2007yk}.
However, most of the branching ratios in the early works are not
predicted precisely, which are usually one or two order smaller than
the experimental data. This may indicate that some contributions,
such as the nonfactorizable and annihilation contributions, are very
important in these decays, which are not included 
in those early works. It has become very urgent to investigate these contributions by
employing proper theoretical models. In Ref. \cite{Cheng:2010yd} the authors
accommodated the experimental data with the QCD factorization (QCDF) approach
\cite{Beneke:QCDF}, which deals with these additional contributions in a very
subtle and technical way. Here we want to adopt yet another theoretical approach,
the perturbative QCD (pQCD) approach \cite{HNLi:pQCD}, which calculate
the nonfactorizable and annihilation contributions in a theoretically systematical
way. These investigations will offer us more detailed knowledge about the
dynamics of $B\to VT$ decays, which is one reason why those decays
are worthy to be studied again.

Another reason why $B\to VT$ decays are meaningful is the
interesting polarization phenomena. In  $B \to VV$ decays, the
transversely polarized contributions of some penguin dominating
channels, such as $B\to (\phi,\rho)K^*$, are nearly the same as the
longitudinal ones \cite{Amsler:2008zzb}. This is quite different from
the prediction of the naive factorization, in which the longitudinal
polarization always dominates. However, in $B\to VT$ decays, such as  $B\to \phi K_2^*$,
the experimental data indicate that the longitudinal polarization
is much larger, while for $B\to \omega K_2^*$  the
longitudinal polarization takes only about a half contribution.
Earlier models such as naive factorization cannot give us any satisfied
explanation. Therefore, employing theoretically more complete models, such as
the QCDF, the pQCD and the soft collinear effective theory, to understand the
phenomena becomes very important.

In a recent paper \cite{Cheng:2010hn}, the authors studied the light cone distribution
amplitudes of the tensor mesons, which make the
calculation of $B\to VT$ decays possible for the QCDF and the pQCD approach.
In their following paper \cite{Cheng:2010yd}, by extracting inputs from the experimental
data they accommodated  $B\to VT$ decays in the frame of the QCDF.
However, some subtle dynamical phenomena is not yet fully
understood, which inspires us to explore these decays under
another approach.
The pQCD approach based on the $k_T$ factorization has
already been used to explore many two body exclusive decays of $B$
meson. The form factors of $B$ to a tensor meson
transition has already been calculated under this
approach \cite{Wang:2010ni}. There are already a few investigations on
the $B$ to a pseudoscalar and a tensor\cite{Zou:2012td} as well as a
charmed meson and a tensor meson
decays \cite{Zou:2012zk,Zou:2012sx,Zou:2012sy}. Though there still
exist few controversies \cite{DescotesGenon:2001hm,Feng:2008zs} on
its feasibility, the predictions based on the pQCD can accommodate
many experimental data well, for example, see Ref. \cite{Li:2004ti}. In
this work, we will put the controversies aside and adopt this
approach to our analysis.

The paper is organized as follows. In
Sec. \ref{section:TheoryFrame}, all the details of the theoretical
frameworks are listed, including the notation conventions, the
Hamiltonian, the kinematics definitions, the wave functions which
are used as the inputs in the pQCD approach, and the analytic formulas
for the Feynman diagrams in the pQCD approach. The numerical results
and discussions are given in Sec. \ref{section:Ndata}, and the last
section is for the summary. In appendix \ref{appendix:forHD} we collect
the expressions of common pQCD functions.

\section{Theoretical details}
\label{section:TheoryFrame}
\subsection{Hamiltonian and kinematics} \label{section:Hamiltonian}

We start from the common low energy effective hamiltonian used in $B$
physics calculations, which are given \cite{Buchalla:1995vs} as
 \begin{eqnarray}
 {\cal H}_{eff} &=& \frac{G_{F}}{\sqrt{2}}
     \bigg\{ \sum\limits_{q=u,c} V_{qb} V_{qD}^{*} \big[
     C_{1}({\mu}) O^{q}_{1}({\mu})
  +  C_{2}({\mu}) O^{q}_{2}({\mu})\Big]\nonumber\\
  &&-V_{tb} V_{tD}^{*} \Big[{\sum\limits_{i=3}^{10}} C_{i}({\mu}) O_{i}({\mu})
  \big ] \bigg\} + \mbox{H.c.} ,
 \label{eq:hamiltonian}
\end{eqnarray}
where $D=s,d$ stands for a down type light quark, $V_{qb(D)}$ and
$V_{tb(D)}$ are Cabibbo-Kobayashi-Maskawa (CKM) matrix elements.
Functions $O_{i}$ ($i=1,...,10$) are local four-quark operators or
the moment type operators:
 \begin{itemize}
 \item  current--current (tree) operators
    \begin{eqnarray}
  O^{q}_{1}=({\bar{q}}_{\alpha}b_{\beta} )_{V-A}
               ({\bar{D}}_{\beta} q_{\alpha})_{V-A},
    \ \ \ \ \ \ \ \ \
   O^{q}_{2}=({\bar{q}}_{\alpha}b_{\alpha})_{V-A}
               ({\bar{D}}_{\beta} q_{\beta} )_{V-A},
    \label{eq:operator12}
    \end{eqnarray}
     \item  QCD penguin operators
    \begin{eqnarray}
      O_{3}=({\bar{D}}_{\alpha}b_{\alpha})_{V-A}\sum\limits_{q^{\prime}}
           ({\bar{q}}^{\prime}_{\beta} q^{\prime}_{\beta} )_{V-A},
    \ \ \ \ \ \ \ \ \
    O_{4}=({\bar{D}}_{\beta} b_{\alpha})_{V-A}\sum\limits_{q^{\prime}}
           ({\bar{q}}^{\prime}_{\alpha}q^{\prime}_{\beta} )_{V-A},
    \label{eq:operator34} \\
     \!\!\!\! \!\!\!\! \!\!\!\! \!\!\!\! \!\!\!\! \!\!\!\!
    O_{5}=({\bar{D}}_{\alpha}b_{\alpha})_{V-A}\sum\limits_{q^{\prime}}
           ({\bar{q}}^{\prime}_{\beta} q^{\prime}_{\beta} )_{V+A},
    \ \ \ \ \ \ \ \ \
    O_{6}=({\bar{D}}_{\beta} b_{\alpha})_{V-A}\sum\limits_{q^{\prime}}
           ({\bar{q}}^{\prime}_{\alpha}q^{\prime}_{\beta} )_{V+A},
    \label{eq:operator56}
    \end{eqnarray}
 \item electro-weak penguin operators
    \begin{eqnarray}
     O_{7}=\frac{3}{2}({\bar{D}}_{\alpha}b_{\alpha})_{V-A}
           \sum\limits_{q^{\prime}}e_{q^{\prime}}
           ({\bar{q}}^{\prime}_{\beta} q^{\prime}_{\beta} )_{V+A},
    \ \ \ \
    O_{8}=\frac{3}{2}({\bar{D}}_{\beta} b_{\alpha})_{V-A}
           \sum\limits_{q^{\prime}}e_{q^{\prime}}
           ({\bar{q}}^{\prime}_{\alpha}q^{\prime}_{\beta} )_{V+A},
    \label{eq:operator78} \\
     O_{9}=\frac{3}{2}({\bar{D}}_{\alpha}b_{\alpha})_{V-A}
           \sum\limits_{q^{\prime}}e_{q^{\prime}}
           ({\bar{q}}^{\prime}_{\beta} q^{\prime}_{\beta} )_{V-A},
    \ \ \ \
    O_{10}=\frac{3}{2}({\bar{D}}_{\beta} b_{\alpha})_{V-A}
           \sum\limits_{q^{\prime}}e_{q^{\prime}}
           ({\bar{q}}^{\prime}_{\alpha}q^{\prime}_{\beta} )_{V-A},
    \label{eq:operator9x}
    \end{eqnarray}
\end{itemize}
 where $\alpha$ and $\beta$ are color indices and $q^\prime$ are
the active quarks at the scale $m_b$, i.e. $q^\prime=(u,d,s,c,b)$.
At the tree level, the operators $O_{7\gamma}$ and $O_{8g}$ do not
contribute, thus they are not listed here. The left handed current
is defined as $({\bar{q}}^{\prime}_{\alpha} q^{\prime}_{\beta}
)_{V-A}= {\bar{q}}^{\prime}_{\alpha} \gamma_\nu (1-\gamma_5)
q^{\prime}_{\beta}  $ and the right handed current
$({\bar{q}}^{\prime}_{\alpha} q^{\prime}_{\beta} )_{V+A}=
{\bar{q}}^{\prime}_{\alpha} \gamma_\nu (1+\gamma_5)
q^{\prime}_{\beta}$. The projection operators are defined as
$P_{L}=(1-\gamma_5)/2$ and $P_{R}=(1+\gamma_5)/2$.
 The combinations $a_i$ of Wilson coefficients are
defined as usual~\cite{Ali:1998eb}:
\begin{eqnarray}
a_1= C_2+C_1/3, &~a_2= C_1+C_2/3, &~ a_3= C_3+C_4/3,  ~a_4=
C_4+C_3/3,~a_5= C_5+C_6/3,\nonumber \\
a_6= C_6+C_5/3, &~a_7= C_7+C_8/3, &~a_8= C_8+C_7/3,~a_9=
C_9+C_{10}/3,
 ~a_{10}= C_{10}+C_{9}/3.
\end{eqnarray}

The calculation is carried out in the rest frame of $B$ meson, the
momenta of $B$ meson ($p_B$), tensor meson ($p_2$) and vector meson
($p_3$) are defined in the light cone coordinates as
\begin{eqnarray}
 p_B=\frac{m_B}{\sqrt{2}}(1,1,{\bf 0_T})\;,\;
 p_2=\frac{m_B}{\sqrt{2}}(1,r_2^2,{\bf 0_T})\;,\;
 p_3=\frac{m_B}{\sqrt{2}}(r_3^2,1,{\bf 0_T})\;\label{eq:momenta}
\end{eqnarray}
with $r_2=m_T/m_B$ and $r_3=m_V/m_B$. In the calculation of the pQCD,
the momenta of the quarks are also related, and they are defined as
follows:
\begin{eqnarray}
 k_1=(0,x_1\frac{m_B}{\sqrt{2}},{\bf k_{1T}})\;,\;
 k_2=(x_2\frac{m_B}{\sqrt{2}},0,{\bf k_{2T}})\;,\;
 k_3=(0,x_3\frac{m_B}{\sqrt{2}},{\bf k_{3T}})\;,\;
\end{eqnarray}
where $k_{1,2,3}$ are the momenta of the light anti-quark in $B$
meson, quarks in tensor and vector mesons, respectively.

\subsection{Wave functions} \label{section:WF}

\subsubsection{$B$ meson}

The $B_{(s)}$ meson wave functions are
  decomposed into the following Lorentz structures:
 \begin{eqnarray}
 &&\int\frac{d^4z}{(2\pi)^4}e^{ik_1\cdot z}\langle0|\bar
 b_{\alpha}(0)d_{\beta}(z)|B_{(s)}(P_1)\rangle\nonumber\\
 &=&\frac{i}{\sqrt{2N_c}}\left\{(\not P_1+m_{B_{(s)}})\gamma_5[\phi_{B_{(s)}}(k_1)-\frac{\not n-\not v}{\sqrt{2}}
 \bar\phi_{B_{(s)}}(k_1)]\right\}_{\beta\alpha},
 \end{eqnarray}
where $\phi_{B_{(s)}}(k_1)$ and $\bar\phi_{B_{(s)}}(k_1)$ are the
leading twist distribution amplitudes. After neglecting the
numerically small contribution term $\bar\phi_{B_{(s)}}(k_1)$
~\cite{Lu:2002ny}, the expression for $\Phi_{B_{(s)}}$ in the
momentum space becomes
\begin{eqnarray}
 \Phi_{B_{(s)}}=\frac{i}{\sqrt{2N_c}}{(\not{P_1}+m_{B_{(s)}})\gamma_5\phi_{B_{(s)}}(k_1)}.
\end{eqnarray}
The calculation of the pQCD is always carried out in the $b$-space, in
which we adopt the following model function
\begin{eqnarray}
 \phi_{B_{(s)}}(x,b)&=&N_{B_{(s)}}x^2(1-x)^2\exp\left[-\frac{1}{2}(\frac{xm_{B_{(s)}}}
 {\omega_b})^2-\frac{\omega_b^2b^2}{2}\right],\label{eq:Bwave}
\end{eqnarray}
where $b$ is the conjugate space coordinate of
$\textbf{k}_{1 T}$. $N_{B_{(s)}}$ is the normalization constant,
which is determined by the normalization condition
\begin{eqnarray}
\int^1_0 dx\phi_{B_{(s)}}(x,b=0)=\frac{f_{B_{(s)}}}{2\sqrt{2N_c}}.
\end{eqnarray}
For $B^{\pm}$ and $B_d^0$ decays, we adopt the value
$\omega_b=0.40~\rm{GeV}$~\cite{Bauer:1988fx}, which is supported by
intensive pQCD studies~\cite{pqcd}. For $B_s$ meson, we will follow
the authors in Ref.~\cite{Ali:2007ff} and adopt the value
$\omega_{b_s}=(0.50\pm0.05)~\rm{GeV}$.

\subsubsection{Vector meson}
 \begin{table}
 \centering
 \caption{The decay constants of vector mesons (in MeV)}
 \begin{tabular}{cccccccc}
 \hline\hline
 \ \ \ $f_{\rho}$  &$f_{K^*}$  &$f_{\omega}$  &$f_{\phi}$  &$f_{\rho}^T$  &$f_{K^*}^T$  &$f_{\omega}^T$  &$f_{\phi}^T$\\
 \hline
 \ \ \ $209\pm2$   &$217\pm5$  &$195\pm3$     &$231\pm4$   &$165\pm9$     &$185\pm10$   &$151\pm9$       &$186\pm9$\\
 \hline
 \end{tabular}\label{vector_decay_constants}
 \end{table}

The decay constants of the vector mesons are defined by
 \begin{equation}
 \langle 0|\bar q_1\gamma_\mu
 q_2|V(p_3,\epsilon)\rangle=f_Vm_V\epsilon_\mu,\;\;\; \langle 0|\bar
 q_1\sigma_{\mu\nu}q_2|V(p_3,\epsilon)\rangle =if^T_V(\epsilon_\mu
 P_{3\nu}-\epsilon_\nu P_{3\mu}).
 \end{equation}
The longitudinal decay constants of the charged mesons can be
extracted experimentally from $\tau^-$ decays and those of the
neutral ones can be extracted from their $e^+e^-$
decays~\cite{Ball:2006eu}, whereas, the transverse decay constants
can be calculated by the QCD sum rules~\cite{Ball:2005vx}. All the
constants for the vector mesons in this paper are collected in
Table~\ref{vector_decay_constants}.

 Up to twist-3 the distribution amplitudes of the light vector mesons
 are summarized as
 \begin{eqnarray}
 \langle V(p_3,\epsilon^*_L)|q_{1\alpha}(0)\bar
 q_{2\beta}(z)|0\rangle&=&-\frac{1}{\sqrt{2N_C}}
 \int_0^1dxe^{ixp_3\cdot z}\left[m_V\not\epsilon^*_L\phi_V(x)+\not\epsilon^*_L\not
 p_3\phi_V^t(x)+m_V\phi_V^s(x)\right]_{\alpha\beta},\nonumber\\
 \langle V(p_3,\epsilon^*_T)|q_{1\alpha}(0)\bar
 q_{2\beta}(z)|0\rangle&=&-\frac{1}{\sqrt{2N_C}}
 \int_0^1dxe^{ixp_3\cdot z}\left[m_V\not\epsilon^*_T\phi_V^v(x)+\not\epsilon^*_T\not
  p_3\phi_V^T(x)\right.\nonumber\\
  &&\left.+m_Vi\epsilon_{\mu\nu\rho\sigma}\gamma_5\gamma^\mu\epsilon^{*\nu}_Tn^\rho v^\sigma
  \phi_V^a(x)\right]_{\alpha\beta},
 \end{eqnarray}
where $x$ is the momentum fraction of the $q_2$ quark. Here $n$ is
the light cone direction along which the meson moves and $v$ is the
opposite direction. With $t=2x-1$ the expression for the twist-2
distribution amplitudes are given by
\begin{eqnarray}
\phi_V(x)=\frac{3f_V}{\sqrt{2N_C}}x(1-x)\left[1+a^\parallel_1C_1^{3/2}(t)
                            +a^\parallel_2C_2^{3/2}(t)\right],\nonumber\\
\phi_V^T(x)=\frac{3f_V}{\sqrt{2N_C}}x(1-x)\left[1+a^\perp_1C_1^{3/2}(t)
                            +a^\perp_2C_2^{3/2}(t)\right].\label{vwavef1}
\end{eqnarray}
 and the corresponding values of the Gegenbauer moments are \cite{vdas}:
 \begin{eqnarray}
 a_{2\rho}^\parallel=a_{2\omega}^\parallel=0.15\pm0.07\;,\;a_{1K^*}^\parallel=0.03\pm0.02\;,\;
 a_{2K^*}^\parallel=0.11\pm0.09\;,\;a_{2\phi}^\parallel=0.18\pm0.08\;,\;\nonumber\\
 a_{2\rho}^\perp=a_{2\omega}^\perp=0.14\pm0.06\;,\;a_{1K^*}^\perp=0.04\pm0.03\;,\;
 a_{2K^*}^\perp=0.10\pm0.08\;,\;a_{2\phi}^\perp=0.14\pm0.07\;.\;
 \end{eqnarray}
We adopt the asymptotic form for the twist-3 distribution
amplitudes:
 \begin{eqnarray}
 &\phi_V^t(x) = \frac{3f_V^T}{2\sqrt{6}} t^2\;,\;&\phi_V^s(x)=\frac{3f_V^T}{2\sqrt{6}}(-t)\;,\nonumber\\
 &\phi_V^v(x) = \frac{3f_V}{8\sqrt{6}} (1+t^2)\;,\;&\phi_V^a(x)=\frac{3f_V}{4\sqrt{6}}(-t)\;.\;\label{vwavef2}
 \end{eqnarray}
%
\subsubsection{Tensor meson}

In the quark model, the tensor meson with $J^{PC}=2^{++}$ has the
angular momentum $L=1$ and spin $S=1$. The ground $SU(3)$ nonet
states are consist of $a_2(1320)$, $f_2(1270)$,
$f_2^{\prime}(1525)$, and $K_2^*(1430)$. Mixing exists for the
$f_2(1270)$ and $f_2^{\prime}(1525)$, just as the $\eta$ and
$\eta^{\prime}$ mixing, and their wave functions can be expressed as
\begin{eqnarray}
f_2&=&f^q \cos\theta_{f_2}+f^s\sin\theta_{f_2},\nonumber\\
f_2^{\prime}&=&f^q \cos\theta_{f_2}-f^s\sin\theta_{f_2},
\end{eqnarray}
where $f^q=\frac{1}{\sqrt{2}}(u\bar u + d\bar d)$ and $f^s=s\bar s$.
The mixing angle $\theta_{f_2}$ is found to be very small,
$\theta_{f_2}=7.8^{\circ}$ \cite{Amsler:2008zzb} and
$\theta_{f_2}=(9\pm1)^{\circ}$ \cite{mixing2}. Therefore, $f_2$ is
nearly an $f^q$ state and $f_2^{\prime}$ is mainly $f^s$.

The spin-2 polarization tensor, which is symmetric and traceless,
satisfies $\epsilon^{\mu\nu}p_{2\nu}=0$ and can be constructed by
spin-1 polarization vectors $\epsilon$ by
\begin{eqnarray}
 \epsilon_{\mu\nu}(\pm2)&=&
 \epsilon_\mu(\pm)\epsilon_\nu(\pm),\;\;\;\;\nonumber\\
 \epsilon_{\mu\nu}(\pm1)&=&\frac{1}{\sqrt2}
 [\epsilon_{\mu}(\pm)\epsilon_\nu(0)+\epsilon_{\nu}(\pm)\epsilon_\mu(0)],\nonumber\\
 \epsilon_{\mu\nu}(0)&=&\frac{1}{\sqrt6}
 [\epsilon_{\mu}(+)\epsilon_\nu(-)+\epsilon_{\nu}(+)\epsilon_\mu(-)]
 +\sqrt{\frac{2}{3}}\epsilon_{\mu}(0)\epsilon_\nu(0).\label{eq:tesorPL}
\end{eqnarray}
In the case that the tensor meson is moving along the $z$-axis, the
polarizations $\epsilon$ can be defined as
\begin{eqnarray}
\epsilon(0)=(|p_2|,0,0,E_2)/m_T\;\;,\;\;\epsilon(\pm1)=(0,\mp1,i,0)/\sqrt{2},
\end{eqnarray}
with $E_2$ as the energy of the tensor meson. Associating with the
tensor momentum defined in Eq. (\ref{eq:momenta}), the polarization
vectors are given in the light cone coordinates by
\begin{eqnarray}
\epsilon(0)=(1,-r_2^2,{\bf{0_T}})/(\sqrt{2}r_2)\;\;,\;\;\epsilon(\pm1)=(0,0,\mp1,i,0)/\sqrt{2}.\label{eq:PLexpression}
\end{eqnarray}

The decay constants of the tensor mesons are defined as
\begin{eqnarray}
 \langle T(p_2)|j_{\mu\nu}(0)|0\rangle&=&f_T m_T^2 \epsilon^*_{\mu\nu},\nonumber\\
 \langle T(p_2)|j_{\mu\nu\rho}(0)|0\rangle&=&-if_T^T
 m_T\left(\epsilon^*_{\mu\delta}p_{2\nu}-\epsilon^*_{\nu\delta}p_{2\mu}\right),
\end{eqnarray}
where the currents are defined as
\begin{eqnarray}
 j_{\mu\nu}(0)&=&\frac{1}{2}[\bar
 q_1(0)\gamma_{\mu}i{\buildrel\leftrightarrow\over D}_{\nu}q_2(0)+\bar
 q_1(0)\gamma_{\nu}i{\buildrel\leftrightarrow\over
 D}_{\mu}q_2(0)],\nonumber\\
 j_{\mu\nu\rho}^{\dagger}(0)&=&\bar
 q_2(0)\sigma_{\mu\nu}i{\buildrel\leftrightarrow\over
 D}_{\rho}q_1(0)
\end{eqnarray}
with ${\buildrel\leftrightarrow\over
D}_{\mu}={\buildrel\rightarrow\over D}_{\mu} -
{\buildrel\leftarrow\over D}_{\mu}$, ${\buildrel\rightarrow\over
D}_{\mu}={\buildrel\rightarrow\over\partial}_{\mu}+ig_s
A_{\mu}^a\lambda^a/2$ and ${\buildrel\leftarrow\over
D}_{\mu}={\buildrel\leftarrow\over\partial}_{\mu}-ig_s
A_{\mu}^a\lambda^a/2$, respectively. These decay constants have
already been studied~\cite{Aliev:1981ju,Aliev:1982ab,Aliev:2009nn}
and we use the recently updated ones with the QCD sum
rules~\cite{Cheng:2010hn}, which are summarized in
Table~\ref{Table:Tdecayconstant}.

\begin{table}[h]
\centering
\caption{Decay constants (in unit of MeV) of tensor mesons from
Ref.~\cite{Cheng:2010hn}.}
\begin{tabular}{cccccccccc}
\hline\hline
   $f_{a_2} $   & $ f_{a_2}^T $
 & $ f_{K_2^*} $ & $ f_{K_2^*}^T $  & $f_{f_2(1270)} $ & $f_{f_2(1270)}^T$
   & $f_{f_2'(1525)} $ & $f_{f_2'(1525)}^T$ \\
\ \ \
   $107\pm6$ & $105\pm 21$    & $ 118\pm 5$  & $77\pm 14$
\ \ \
 & $102\pm 6$ & $117\pm25$    & $126\pm 4$  & $65\pm 12$\\
\hline \hline
\end{tabular}\label{Table:Tdecayconstant}
 \end{table}
The light cone distribution amplitudes (LCDAs) of the tensor mesons are also recently studied by
Ref.~\cite{Cheng:2010hn} and we follow the notations in
Ref.~\cite{Wang:2010ni} to summarize them up to twist-3 as
\begin{eqnarray}
\langle T(p_2,\epsilon)|q_{1\alpha} (0)\bar q_{2\beta}(z) |0\rangle
&=&\frac{1}{\sqrt{2N_c}}\int_0^1 dx e^{ixp_2\cdot z} \left[m_T\not\!
\epsilon^*_{\bullet L} \phi_T(x) +\not\! \epsilon^*_{\bullet
L}\not\! p_2 \phi_{T}^{t}(x) +m_T^2\frac{\epsilon_{\bullet} \cdot
v}{p_2\cdot v} \phi_T^s(x)\right]_{\alpha\beta},
\label{eq:lpwf}\\
 \langle T(p_2,\epsilon)|q_{1\alpha}
(0) \bar q_{2\beta}(z) |0\rangle &=&\frac{1}{\sqrt{2N_c}}\int_0^1 dx
e^{ixp_2\cdot z} \left[m_T\not\! \epsilon^*_{\bullet T}\phi_T^v(x)+
\not\!\epsilon^*_{\bullet T}\not\! p_2\phi_T^T(x)+m_T
i\epsilon_{\mu\nu\rho\sigma}\gamma_5\gamma^\mu\right.\nonumber\\
&&\left.\times\epsilon_{\bullet
T}^{*\nu} n^\rho v^\sigma \phi_T^a(x)\right]_{\alpha\beta}\;,
\label{eq:tpwf}
\end{eqnarray}
with $\epsilon^{0123}=1$ adopted. Eq.~(\ref{eq:lpwf}) is for the
longitudinal polarized mesons ($h=0$) and Eq.~(\ref{eq:tpwf}) for
the transverse polarized ones ($h=\pm 1$). $x$ is the momentum
fraction associated with the $q_2$ quark. $n$ is the light cone
direction along with tensor meson moves and $v$ is the opposite
direction. $\epsilon_{\bullet}$ is defined by
\begin{equation}
\epsilon_{\bullet\mu}\equiv\frac{\epsilon_{\mu\nu} v^\nu}{p_2\cdot
v}m_T.\label{eq:epsilondot}
\end{equation}
With the momenta and polarizations defined in the above paragraphs,
Eq. (\ref{eq:epsilondot}) can be reexpressed by
\begin{equation}
\epsilon_{\bullet\mu}=\frac{2m_T}{m_B^2}\epsilon_{\mu\nu}p_B^{\nu}
\end{equation}
up to the leading power of $r_2$. We follow the symbols in
Ref.~\cite{Wang:2010ni}, and list the expressions of LCDAs as
\begin{eqnarray}
&&\phi_{T}(x)=\frac{f_{T}}{2\sqrt{2N_c}}\phi_{||}(x),\;\;\;
\phi_{T}^t(x)=\frac{f_{T}^T}{2\sqrt{2N_c}}h_{||}^{(t)}(x),\nonumber\\
&&\phi_{T}^s(x)=\frac{f_{T}^T}{4\sqrt{2N_c}}
\frac{d}{dx}h_{||}^{(s)}(x),\hspace{3mm}
\phi_{T}^T(x)=\frac{f_{T}^T}{2\sqrt{2N_c}}\phi_{\perp}(x)
,\nonumber\\
&&\phi_{T}^v(x)=\frac{f_{T}}{2\sqrt{2N_c}}g_{\perp}^{(v)}(x),
\hspace{3mm}\phi_{T}^a(x)=\frac{f_{T}}{8\sqrt{2N_c}}
\frac{d}{dx}g_{\perp}^{(a)}(x).
\end{eqnarray}
The twist-2 LCDAs can be expanded in terms of the Gegenbauer
polynomials, and their asymptotic form are given by
\begin{eqnarray}
\phi_{\parallel,\perp}(x)=30x(1-x)(2x-1)
\end{eqnarray}
with the normalization conditions
\begin{equation}
\int_0^1dx(2x-1)\phi_{\parallel,\perp}(x)=1.
\end{equation}
By using the QCD equations of motion, the twist-3 two partons distribution amplitudes (DAs)
can be related to the twist-2 ones and the tree partons
DAs~\cite{Ball:1998ff,Ball:1998sk}. Their expressions for the
asymptotic forms are given by~\cite{Cheng:2010hn}
\begin{eqnarray}
h_\parallel^{(t)}(x) & = & \frac{15}{2}(2x-1)(1-6x+6x^2) ,\;\;\;
h_{||}^{(s)}(x)   = 15x(1-x)(2x-1),\\
g_\perp^{(a)}(x) & = & 20x(1-x)(2x-1) ,\;\;\; g_\perp^{(v)}(x)
=5(2x-1)^3.
\end{eqnarray}

\subsection{Analytic formulae} \label{section:Aformula}

In this subsection, we list the pQCD formulas for all the possible
Feynman diagrams. In the diagrams we use $M_{2,3}$ to denote the
tensor and vector mesons, respectively. At the tree level, the
Feynman diagrams in the pQCD can be divided into two types according to their typological structures: the emission diagrams, in which the
light quark in $B$ meson enter one of the light mesons as a
spectator, and the annihilation diagrams, in which both of the two
quarks in $B$ mesons are absorbed by the electro-weak operator.
According to the polarizations, we can list the formulas in two
parts, the longitudinal polarizations and the transverse ones.
For simplicity we only list the
amplitude functions for the longitudinal ones. The
transverse polarized ones can be calculated in the same way with the
corresponding wave functions.
\begin{figure}[h]
\centering
\includegraphics[scale=0.5]{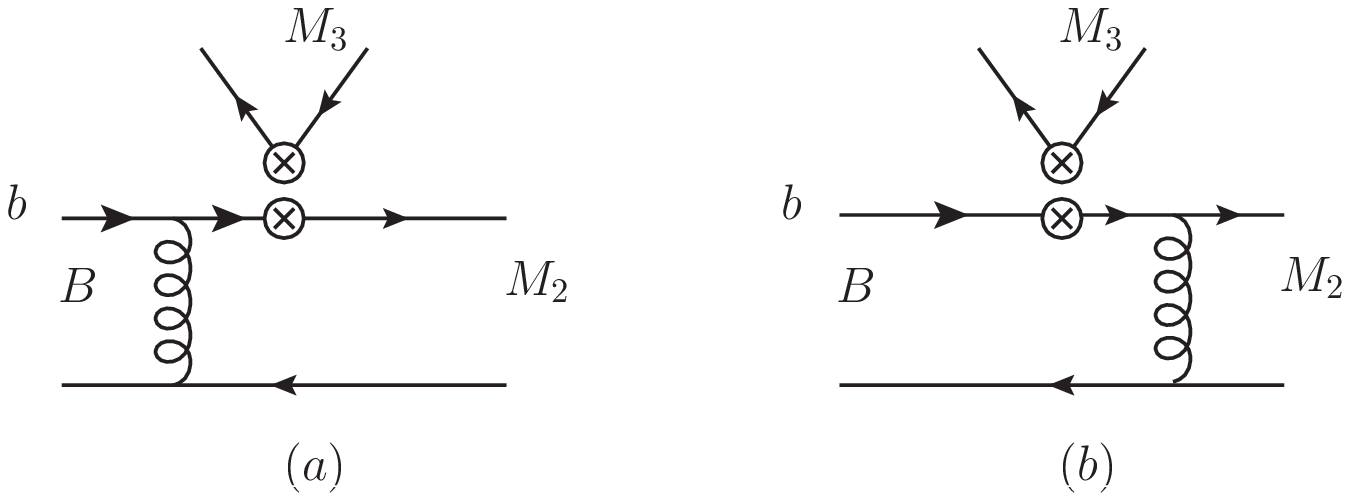}
\hspace{1cm}
\includegraphics[scale=0.5]{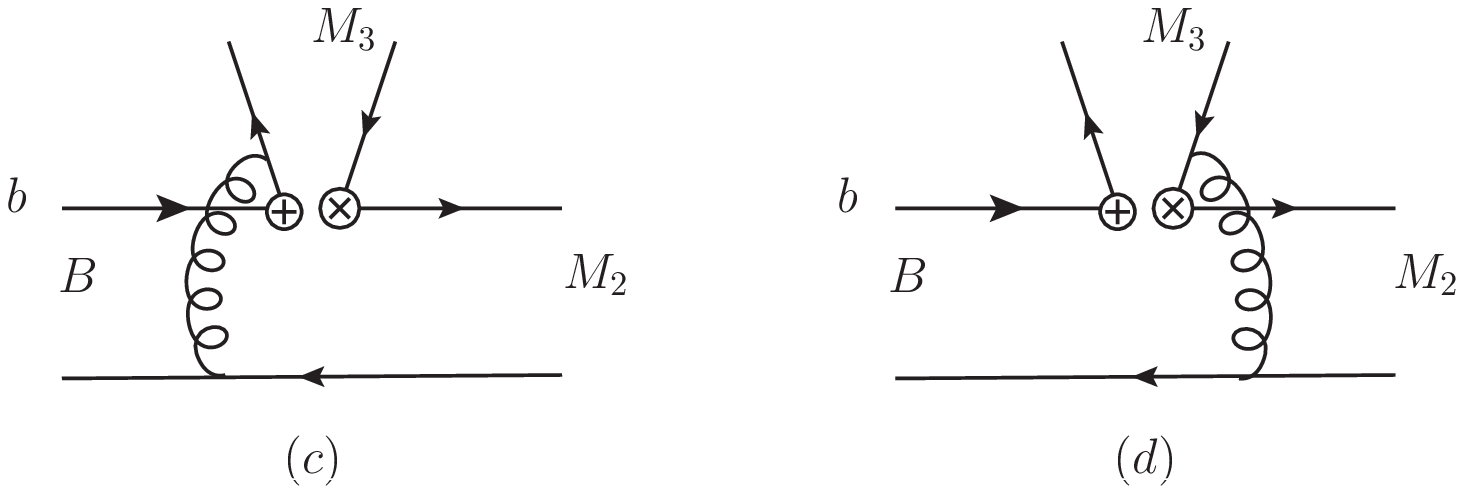}
\caption{The emission diagrams with a vector meson emitted.}
\label{fig:m3e}
\end{figure}
The factorizable emission diagrams are shown as the first two
diagrams in Fig.~\ref{fig:m3e}. Since the tensor meson can not be
generated from the vector or axial vector current, only the vector
meson can be emitted. The expressions for all possible Lorentz
structures are given as follows.

{\small
\begin{itemize}
\item (V-A)(V-A) factorizable emission diagrams:
\begin{eqnarray}
F_{vef}^{LL}(a_i)&=&8\sqrt{\frac{2}{3}}\pi m_B^4 f_V C_F\int_0^1
dx_1
dx_2 \int_0^{1/\Lambda_{QCD}}b_1 db_1 b_2 db_2 \phi_B(x_1)\nonumber\\
&&\left\{\left(r_2(2x_2-1)(\phi_T^t(x_2)-\phi_T^s(x_2))+(2-x_2)\phi_T(x_2)\right)a_i(t_{vef}^1)E_e(t_{vef}^1)
\right.\nonumber\\
&&\left. \times h_e(\sqrt{|\alpha_{ef1}^2|},\sqrt{|\beta_{ef1}^2|},
b_2,b_1)S_t(x_2)-2r_2\phi_T^s(x_2)a_i(t_{vef}^2)E_e(t_{vef}^2)\right.\nonumber\\
&&\left.\times
h_e(\sqrt{|\alpha_{ef2}^2|},\sqrt{|\beta_{ef2}^2|},b_1,b_2)
S_t(x_1)\right\}\;,
\end{eqnarray}
where the explicit expressions of scales $t_{vef}^{1,2}$, the
production of coupling $\alpha_s$ and Sudakov factor
$E_e(t_{vef}^{1,2})$, the function of hard kernel $h_e$, the
parameters $\alpha_{(ef1,ef2)}^2$ and $\beta_{(ef1,ef2)}^2$, and the jet
function $S_t(x)$ are all collected in Appendix
\ref{appendix:forHD}. In the following analytic formulas, the
expressions of all the additional functions can also be found in the same
appendix.
\item (V-A)(V+A) factorizable emission diagrams:
\begin{equation}
F_{vef}^{LR}(a_i)=F_{vef}^{LL}(a_i)\;,
\end{equation}
\item (S-P)(S+P) factorizable emission diagrams:
\begin{equation}
F_{vef}^{SP}(a_i)=0\;.
\end{equation}
\end{itemize}
}
\begin{figure}[h]
\centering
\includegraphics[scale=0.5]{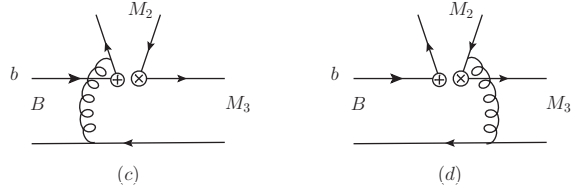}
\caption{The nonfactorizable emission diagrams with a tensor meson
emitted.}\label{fig:m2e}
\end{figure}
There are two possible types of nonfactorizable emission diagrams,
one has the vector meson emitted and the other has the tensor meson
emitted. They are depicted by the last two diagrams of
Fig.~\ref{fig:m3e} and Fig.~\ref{fig:m2e} respectively. We use the
index $ten$ to represent the tensor meson emission and $ven$ for
vector meson emission. The expressions are given by:{\small
\begin{itemize}
\item (V-A)(V-A) nonfactorizable emission diagrams with vector meson emission:
\begin{eqnarray}
F_{ven}^{LL}(a_i)&=&\frac{32}{3}\pi m_B^4 C_F \int_0^1 dx_1 dx_2
dx_3
\int_0^{1/\Lambda_{QCD}}b_1db_1 b_3 db_3 \phi_B(x_1)\phi_V(x_3)\nonumber\\
&&\left\{\left(r_2(1-x_2)(\phi_T^s(x_2)+\phi_T^t(x_2))+x_3\phi_T(x_2)\right)a_i(t_{ven}^1)E_{en}(t_{ven}^1,1,2)\right.\nonumber\\
&&\left. \times h_{en}(\sqrt{|\alpha_{en1}^2|},\sqrt{|\beta_{en1}^2|},
b_1,b_3)+\left(r_2(x_2-1)(\phi_T^s(x_2)-\phi_T^t(x_2))+(x_2+x_3-2)\phi_T(x_2))\right)\right.\nonumber\\
&&\left. \times a_i(t_{ven}^2)E_{en}(t_{ven}^2,1,2)
 h_{en}(\sqrt{|\alpha_{en2}^2|},\sqrt{|\beta_{en2}^2|},b_1,b_3)\right\}\;,
\end{eqnarray}
\item (V-A)(V+A) nonfactorizable emission diagrams with vector meson emission:
\begin{eqnarray}
F_{ven}^{LR}(a_i)&=&\frac{32}{3}\pi m_B^4 C_F r_3 \int_0^1 dx_1 dx_2
dx_3 \int_0^{1/\Lambda_{QCD}}b_1db_1 b_3 db_3
\phi_B(x_1)\nonumber\\
&&\left\{\left(r_2((x_2-1)(\phi_T^s(x_2)-\phi_T^t(x_2))(\phi_V^s(x_3)+\phi_V^t(x_3))-x_3(\phi_T^s(x_2)+\phi_T^t(x_2))(\phi_V^s(x_3)-\phi_V^t(x_3)))\right.\right.\nonumber\\
&&\left.\left.+x_3\phi_T(x_2)(\phi_V^s(x_3)-\phi_V^t(x_3))\right)a_i(t_{ven}^1)E_{en}(t_{ven}^1,1,2)h_{en}(\sqrt{|\alpha_{en1}^2|},\sqrt{|\beta_{en1}^2|},b_1,b_3)\right.\nonumber\\
&&\left.-\left(r_2(x_2(\phi_V^s(x_3)-\phi_V^t(x_3))(\phi_T^s(x_2)-\phi_T^t(x_2))+x_3(\phi_V^s(x_3)+\phi_V^t(x_3))(\phi_T^s(x_2)+\phi_T^t(x_2))\right.\right.\nonumber\\
&&\left.\left.-2(\phi_V^s(x_3)\phi_T^s(x_2)+\phi_V^t(x_3)\phi_T^t(x_2)))-(x_3-1)\phi_T(x_2)(\phi_V^s(x_3)+\phi_V^t(x_3))\right)\right.\nonumber\\
&&\left.\times
a_i(t_{ven}^2)E_{en}(t_{ven}^2,1,2)h_{en}(\sqrt{|\alpha_{en2}^2|},\sqrt{|\beta_{en2}^2|},b_1,b_3)\right\}\;,
\end{eqnarray}
\item (S-P)(S+P) nonfactorizable emission diagrams with vector meson emission:
\begin{eqnarray}
F_{ven}^{SP}(a_i)&=&\frac{32}{3}\pi m_B^4 C_F \int_0^1 dx_1 dx_2
dx_3 \int_0^{1/\Lambda_{QCD}}b_1db_1 b_3 db_3
\phi_B(x_1)\phi_V(x_3)\nonumber\\
&&\left\{\left(r_2(1-x_2)(\phi_T^s(x_2)-\phi_T^t(x_2))+\phi_T(x_2)(-x_2+x_3+1)\right)
a_i(t_{ven}^1)E_{en}(t_{ven}^1,1,2)\right.\nonumber\\
&&\left. \times h_{en}(\sqrt{|\alpha_{en1}^2|},
\sqrt{|\beta_{en1}^2|},b_1,b_3) +\left(r_2(x_2-1)(\phi_T^t(x_2)+\phi_T^s(x_2))+\phi_T(x_2)(x_3-1)\right)a_i(t_{ven}^2)\right.\nonumber\\
&&\left.\times E_{en}(t_{ven}^2,1,2) h_{en}(\sqrt{|\alpha_{en2}^2|},\sqrt{|\beta_{en2}^2|},b_1,b_3)\right\}\;.
\end{eqnarray}
\item (V-A)(V-A) nonfactorizable emission diagrams with tensor meson emission:
\begin{eqnarray}
F_{ten}^{LL}(a_i)&=&\frac{32}{3}\pi m_B^4 C_F \int_0^1 dx_1 dx_2
dx_3 \int_0^{1/\Lambda_{QCD}}b_1db_1 b_2 db_2
\phi_B(x_1)\phi_T(x_2)\nonumber\\
&&\left\{\left(x_2\phi_V(x_3)-r_3(x_3-1)(\phi_V^s(x_3)+\phi_V^t(x_3))\right)a_i(t_{ten}^1)E_{en}(t_{ten}^1,1,3)h_{en}(\sqrt{|\alpha_{en1}^{\prime 2}|},\sqrt{|\beta_{en1}^{\prime 2}|},b_1,b_2)\right.\nonumber\\
&&\left.+\left(r_3(x_3-1)(\phi_V^s(x_3)-\phi_V^t(x_3))+\phi_V(x_3)(x_2+x_3-2)\right)a_i(t_{ten}^2)E_{en}(t_{ten}^2,1,3)\right.\nonumber\\
&&\left. \times h_{en}(\sqrt{|\alpha_{en2}^{\prime
2}|},\sqrt{|\beta_{en2}^{\prime 2}|},b_1,b_2)\right\}\;,
\end{eqnarray}
\item (V-A)(V+A) nonfactorizable emission diagrams with tensor meson emission:
\begin{eqnarray}
F_{ten}^{LR}(a_i)&=&\frac{32}{3}\pi m_B^4 C_F r_2 \int_0^1 dx_1 dx_2
dx_3 \int_0^{1/\Lambda_{QCD}}b_1db_1 b_2 db_2
\phi_B(x_1)\nonumber\\
&&\left\{\left(r_3(x_2(\phi_V^s(x_3)+\phi_V^t(x_3))(\phi_T^t(x_2)-\phi_T^s(x_2))+(1-x_3)(\phi_T^s(x_2)+\phi_T^t(x_2))(\phi_V^t(x_3)-\phi_V^s(x_3)))\right.\right.\nonumber\\
&&\left.\left.+x_2\phi_V(x_3)(\phi_T^s(x_2)-\phi_T^t(x_2))\right)a_i(t_{ten}^1)E_{en}(t_{ten}^1,1,3)h_{en}(\sqrt{|\alpha_{en1}^{\prime 2}|},\sqrt{|\beta_{en1}^{\prime 2}|},b_1,b_2)\right.\nonumber\\
&&\left.+\left(r_3(-x_2(\phi_V^s(x_3)+\phi_V^t(x_3))(\phi_T^s(x_2)+\phi_T^t(x_2))+x_3(\phi_V^s(x_3)-\phi_V^t(x_3))(\phi_T^t(x_2)-\phi_T^s(x_2))\right.\right.\nonumber\\
&&\left.\left.+2(\phi_V^t(x_3)\phi_T^t(x_2)+\phi_V^s(x_3)\phi_T^s(x_2)))+(x_2-1)\phi_V(x_3)(\phi_T^s(x_2)+\phi_T^t(x_2))\right)\right.\nonumber\\
&&\left.\times
a_i(t_{ten}^2)E_{en}(t_{ten}^2,1,3)h_{en}(\sqrt{|\alpha_{en2}^{\prime
2}|},\sqrt{|\beta_{en2}^{\prime 2}|},b_1,b_2)\right\}\;,
\end{eqnarray}
\item (S-P)(S+P) nonfactorizable emission diagrams with tensor meson emission:
\begin{eqnarray}
F_{ten}^{SP}(a_i)&=&\frac{32}{3}\pi m_B^4 C_F \int_0^1 dx_1 dx_2
dx_3 \int_0^{1/\Lambda_{QCD}}b_1db_1 b_2 db_2
\phi_B(x_1)\phi_T(x_2)\nonumber\\
&&\left\{\left(\phi_V(x_3)(x_2-x_3+1)-r_3(x_3-1)(\phi_V^s(x_3)-\phi_V^t(x_3))\right)a_i(t_{ten}^1)E_{en}(t_{ten}^1,1,3)\right.\nonumber\\
&&\left. \times h_{en}(\sqrt{|\alpha_{ef1}^{\prime 2}|},\sqrt{|\beta_{ef1}^{\prime 2}|},b_1,b_2)+\left(r_3(x_3-1)(\phi_V^s(x_3)+\phi_V^t(x_3))+(x_2-1)\phi_V(x_3)\right)a_i(t_{ten}^2)\right.\nonumber\\
&&\left.\times E_{en}(t_{ten}^2,1,3)h_{en}(\sqrt{|\alpha_{en2}^{\prime
2}|},\sqrt{|\beta_{en2}^{\prime 2}|},b_1,b_2)\right\}\;.
\end{eqnarray}
\end{itemize}
}

\begin{figure}[h]
\centering
\includegraphics[scale=0.5]{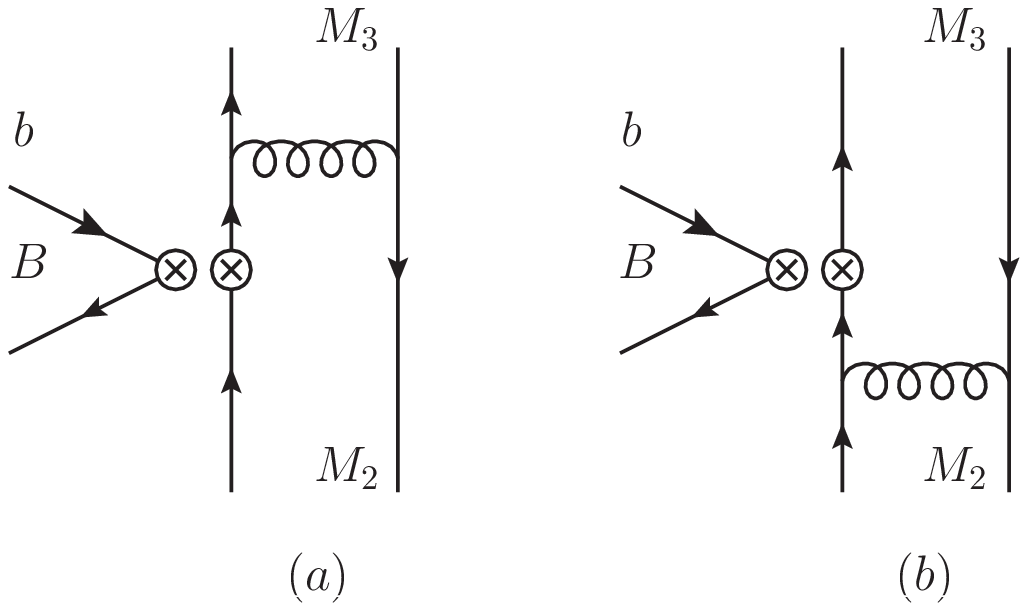}
\hspace{1cm}
\includegraphics[scale=0.5]{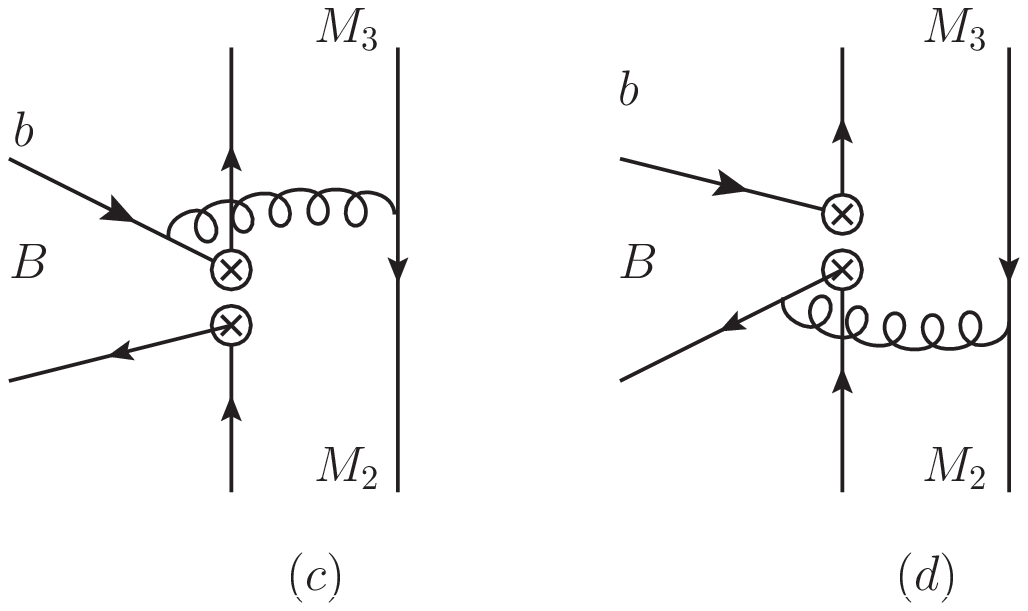}
\caption{The annihilation diagrams with the electro-weak generated
anti-quark entering the tensor meson.}\label{fig:m2a}
\end{figure}

\begin{figure}[h]
\centering
\includegraphics[scale=0.5]{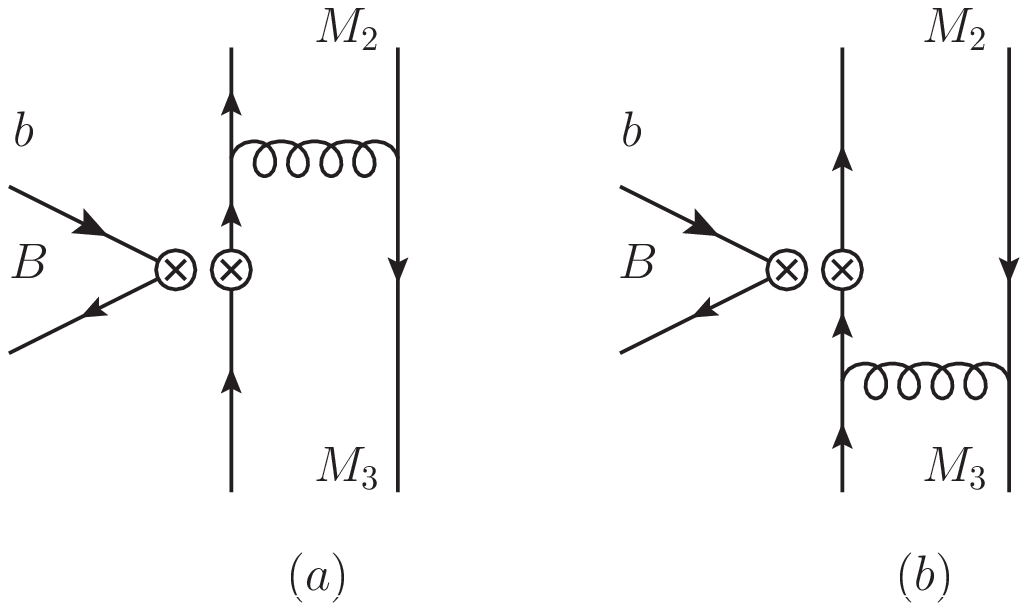}
\hspace{1cm}
\includegraphics[scale=0.5]{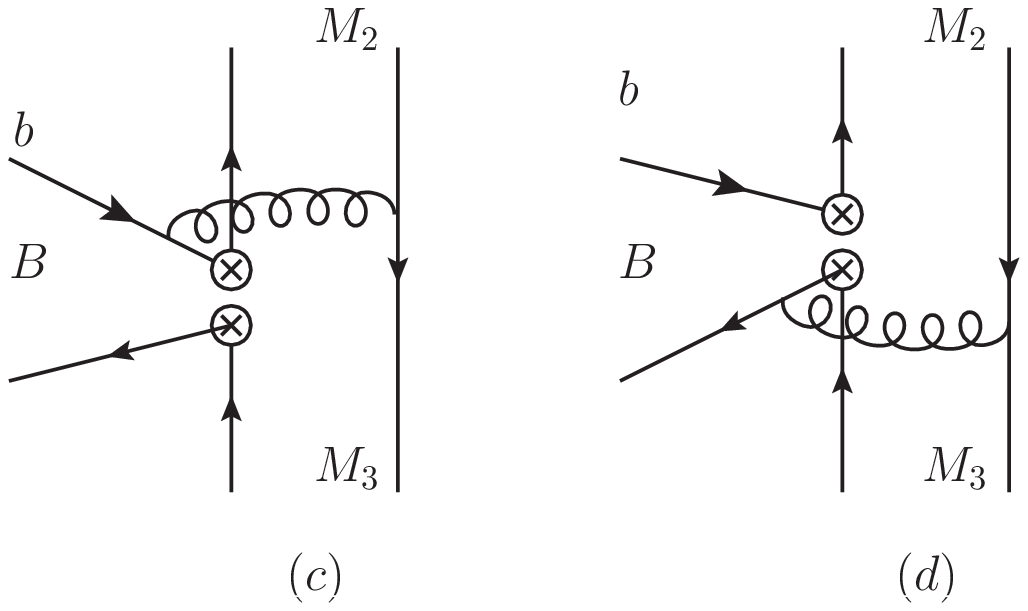}
\caption{The annihilation diagrams with the electro-weak generated
anti-quark entering the vector meson.}\label{fig:m3a}
\end{figure}

According to which meson has the anti-quark generated from the weak
vertex, the annihilation diagrams are also divided into two
types, as depicted in Figs.~\ref{fig:m2a} and \ref{fig:m3a}.
We use the first letter of the index ``$v$" to denote the case that
the quark enters the vector meson and ``$t$" to denote that the quark
enters the tensor meson.
For the factorizable annihilation diagrams, which are the first two
diagrams in Figs.~\ref{fig:m2a} and \ref{fig:m3a}, the corresponding
functions are given as follows.{\small
\begin{itemize}
\item (V-A)(V-A) factorizable annihilation diagrams with the quark entering the vector meson:
\begin{eqnarray}
F_{vaf}^{LL}(a_i)&=&8\sqrt{\frac{2}{3}}\pi m_B^4 f_B C_F\int_0^1
dx_2 dx_3 \int_0^{1/\Lambda_{QCD}}b_2 db_2 b_3
db_3\nonumber\\
&&\left\{\left(-2r_2r_3\phi_V^s(x_3)(-\phi_T^s(x_2)(x_2+1)+\phi_T^t(x_2)(1-x_2))-x_2\phi_T(x_2)\phi_V(x_3)\right)a_i(t_{vaf}^1)\right.\nonumber\\
&&\left.\times E_a(t_{vaf}^1)h_a(\sqrt{|\alpha_{af1}^{2}|},\sqrt{|\beta_{af1}^{2}|},b_2,b_3)S_t(x_2)\right.\nonumber\\
&&\left.+\left(-2r_2r_3\phi_T^s(x_2)((2-x_3)\phi_V^s(x_3)+x_3\phi_V^t(x_3))+(1-x_3)\phi_T(x_2)\phi_V(x_3)\right)a_i(t_{vaf}^2)\right.\nonumber\\
&&\left.\times
E_a(t_{vaf}^2)h_a(\sqrt{|\alpha_{af2}^{2}|},\sqrt{|\beta_{af2}^{2}|},b_3,b_2)S_t(x_3)\right\}\;,
\end{eqnarray}
\item (V-A)(V+A) factorizable annihilation diagrams with the quark entering the vector meson:
\begin{eqnarray}
F_{vaf}^{LR}(a_i)=F_{vaf}^{LL}(a_i)\;,
\end{eqnarray}
\item (S-P)(S+P) factorizable annihilation diagrams with the quark entering the vector meson:
\begin{eqnarray}
F_{vaf}^{SP}(a_i)&=&16\sqrt{\frac{2}{3}}\pi m_B^4 f_B C_F\int_0^1
dx_2 dx_3 \int_0^{1/\Lambda_{QCD}}b_2 db_2 b_3
db_3\nonumber\\
&&\left\{\left(r_2x_2\phi_V(x_3)(\phi_T^t(x_2)-\phi_T^s(x_2))+2r_3\phi_T(x_2)\phi_V^s(x_3)\right)a_i(t_{vaf}^1)E_a(t_{vaf}^1)
\right.\nonumber\\
&&\left.\times h_a(\sqrt{|\alpha_{af1}^{2}|},\sqrt{|\beta_{af1}^{2}|},b_2,b_3)S_t(x_2)+\left(r_3(1-x_3)\phi_T(x_2)(\phi_V^s(x_3)+\phi_V^t(x_3))-2r_2\phi_T^s(x_2)
\phi_V(x_3)\right)\right.\nonumber\\
&&\left.\times a_i(t_{vaf}^2)E_a(t_{vaf}^2)h_a(\sqrt{|\alpha_{af2}^{2}|},\sqrt{|\beta_{af2}^{2}|},b_3,b_2)S_t(x_3)\right\}\;.
\end{eqnarray}
\item (V-A)(V-A) factorizable annihilation diagrams with the quark entering the tensor meson:
\begin{eqnarray}
F_{taf}^{LL}(a_i)&=&8\sqrt{\frac{2}{3}}\pi m_B^4 f_B C_F\int_0^1
dx_2 dx_3 \int_0^{1/\Lambda_{QCD}}b_2 db_2 b_3
db_3\nonumber\\
&&\left\{\left(2r_2r_3\phi_T^s(x_2)((x_3+1)\phi_V^s(x_3)+(x_3-1)\phi_V^t(x_3))-x_3\phi_T(x_2)\phi_V(x_3)\right)a_i(t_{taf}^1)\right.\nonumber\\
&&\left.\times E_a(t_{taf}^1)h_a(\sqrt{|\alpha_{af1}^{\prime 2}|},\sqrt{|\beta_{af1}^{\prime 2}|},b_3,b_2)S_t(x_3)\right.\nonumber\\
&&\left.+\left(2r_2r_3\phi_V^s(x_3)((x_2-2)\phi_T^s(x_2)-x_2\phi_T^t(x_2))+(1-x_2)\phi_T(x_2)\phi_V(x_3)\right)a_i(t_{taf}^2)\right.\nonumber\\
&&\left.\times E_a(t_{taf}^2)h_a(\sqrt{|\alpha_{af2}^{\prime
2}|},\sqrt{|\beta_{af2}^{\prime 2}|},b_2,b_3)S_t(x_2)\right\}\;,
\end{eqnarray}
\item (V-A)(V+A) factorizable annihilation diagrams with the quark entering the tensor meson:
\begin{eqnarray}
F_{taf}^{LR}(a_i)=F_{taf}^{LL}(a_i)\;,
\end{eqnarray}
\item (S-P)(S+P) factorizable annihilation diagrams with the quark entering the tensor meson:
\begin{eqnarray}
F_{taf}^{SP}(a_i)&=&16\sqrt{\frac{2}{3}}\pi m_B^4 f_B C_F\int_0^1
dx_2 dx_3 \int_0^{1/\Lambda_{QCD}}b_2 db_2 b_3
db_3\nonumber\\
&&\left\{\left(r_3x_3\phi_T(x_2)(\phi_V^t(x_3)-\phi_V^s(x_3))+2r_2\phi_T^s(x_2)\phi_V(x_3)\right)a_i(t_{taf}^1)E_a(t_{taf}^1)\right.\nonumber\\
&&\left.\times h_a(\sqrt{|\alpha_{af1}^{\prime 2}|},\sqrt{|\beta_{af1}^{\prime 2}|},b_3,b_2)S_t(x_3)+\left(r_2\phi_V(x_3)(1-x_2)(\phi_T^s(x_2)+\phi_T^t(x_2))
-2r_3\phi_T(x_2)\phi_V^s(x_3)\right)\right.\nonumber\\
&&\left.\times a_i(t_{taf}^2)E_a(t_{taf}^2)h_a(\sqrt{|\alpha_{af2}^{\prime
2}|},\sqrt{|\beta_{af2}^{\prime 2}|},b_2,b_3)S_t(x_2)\right\}\;.
\end{eqnarray}
\end{itemize}}
The nonfactorizable annihilation diagrams are depicted by the last
two diagrams in Figs.~\ref{fig:m2a} and \ref{fig:m3a}, and their
corresponding functions are given in the following. {\small
\begin{itemize}
\item (V-A)(V-A) nonfactorizable annihilation diagrams with the quark entering the vector meson:
\begin{eqnarray}
F_{van}^{LL}(a_i)&=&\frac{32}{3}\pi m_B^4 C_F \int_0^1 dx_1 dx_2
dx_3 \int_0^{1/\Lambda_{QCD}}b_1db_1 b_3 db_3
\phi_B(x_1)\nonumber\\
&&\left\{\left(r_2r_3\phi_T^s(x_2)(\phi_V^s(x_3)(x_2-x_3+3)+\phi_V^t(x_3)(x_2+x_3-1))\right.\right.\nonumber\\
&&\left.\left.-r_2r_3\phi_T^t(x_2)(\phi_V^s(x_3)(x_2+x_3-1)+\phi_V^t(x_3)(x_2-x_3-1))\right.\right.\nonumber\\
&&\left.\left.+(x_3-1)\phi_T(x_2)\phi_V(x_3)\right)a_i(t_{van}^1)E_{an}(t_{van}^1)h_{an}(\sqrt{|\alpha_{an1}^{2}|},\sqrt{|\beta_{an1}^{2}|},b_2,b_1)\right.\nonumber\\
&&\left.+\left(r_2r_3\phi_T^s(x_2)(\phi_V^s(x_3)(-x_2+x_3-1)+\phi_V^t(x_3)(x_2+x_3-1))\right.\right.\nonumber\\
&&\left.\left.-r_2r_3\phi_T^t(x_2)(\phi_V^s(x_3)(x_2+x_3-1)+\phi_V^t(x_3)(-x_2+x_3-1))\right.\right.\nonumber\\
&&\left.\left.+x_2\phi_T(x_2)\phi_V(x_3)\right)a_i(t_{van}^2)E_{an}(t_{van}^2)h_{an}(\sqrt{|\alpha_{an2}^{2}|},\sqrt{|\beta_{an2}^{2}|},b_2,b_1)\right\}\;,
\end{eqnarray}
\item (V-A)(V+A) nonfactorizable annihilation diagrams with the quark entering the vector meson:
\begin{eqnarray}
F_{van}^{LR}(a_i)&=&\frac{32}{3}\pi m_B^4 C_F \int_0^1 dx_1 dx_2
dx_3 \int_0^{1/\Lambda_{QCD}}b_1db_1 b_3 db_3
\phi_B(x_1)\nonumber\\
&&\left\{\left(r_2\phi_V(x_3)(2-x_2)(\phi_T^s(x_2)+\phi_T^t(x_2))+r_3(x_3+1)\phi_T(x_2)(\phi_V^s(x_3)-\phi_V^t(x_3))\right)a_i(t_{van}^1)\right.\nonumber\\
&&\left.\times E_{an}(t_{van}^1)h_{an}(\sqrt{|\alpha_{an1}^{2}|},\sqrt{|\beta_{an1}^{2}|},b_2,b_1)\right.\nonumber\\
&&\left.+\left(r_2x_2\phi_V(x_3)(\phi_T^s(x_2)+\phi_T^t(x_2))+r_3(1-x_3)\phi_T(x_2)(\phi_V^s(x_3)-\phi_V^t(x_3))\right)a_i(t_{van}^2)
\right.\nonumber\\
&&\left.\times
E_{an}(t_{van}^2)h_{an}(\sqrt{|\alpha_{an2}^{2}|},\sqrt{|\beta_{an2}^{2}|},b_2,b_1)\right\}\;,
\end{eqnarray}
\item (S-P)(S+P) nonfactorizable annihilation diagrams with the quark entering the vector meson:
\begin{eqnarray}
F_{van}^{SP}(a_i)&=&\frac{32}{3}\pi m_B^4 C_F \int_0^1 dx_1 dx_2
dx_3 \int_0^{1/\Lambda_{QCD}}b_1db_1 b_3 db_3
\phi_B(x_1)\nonumber\\
&&\left\{\left(r_2r_3\phi_T^s(x_2)(\phi_V^s(x_3)(x_2-x_3+3)-\phi_V^t(x_3)(x_2+x_3-1))\right.\right.\nonumber\\
&&\left.\left.-r_2r_3\phi_T^t(x_2)(\phi_V^t(x_3)(x_2-x_3-1)-\phi_V^s(x_3)(x_2+x_3-1))\right.\right.\nonumber\\
&&\left.\left.-x_2\phi_T(x_2)\phi_V(x_3)\right)a_i(t_{van}^1)E_{an}(t_{van}^1)h_{an}(\sqrt{|\alpha_{an1}^{2}|},\sqrt{|\beta_{an1}^{2}|},b_2,b_1)\right.\nonumber\\
&&\left.+\left(r_2r_3(-\phi_T^s(x_2)(\phi_V^s(x_3)(x_2-x_3+1)+\phi_V^t(x_3)(x_2+x_3-1))\right.\right.\nonumber\\
&&\left.\left.+r_2r_3\phi_T^t(x_2)(\phi_V^t(x_3)(x_2-x_3+1)+\phi_V^s(x_3)(x_2+x_3-1))\right.\right.\nonumber\\
&&\left.\left.-(x_3-1)\phi_T(x_2)\phi_V(x_3)\right)a_i(t_{van}^2)E_{an}(t_{van}^2)h_{an}(\sqrt{|\alpha_{an2}^{2}|},\sqrt{|\beta_{an2}^{2}|},b_2,b_1)\right\}\;.
\end{eqnarray}
\item (V-A)(V-A) nonfactorizable annihilation diagrams with the quark entering the tensor meson:
\begin{eqnarray}
F_{tan}^{LL}(a_i)&=&\frac{32}{3}\pi m_B^4 C_F \int_0^1 dx_1 dx_2
dx_3 \int_0^{1/\Lambda_{QCD}}b_1db_1 b_3 db_3
\phi_B(x_1)\nonumber\\
&&\left\{\left(r_2r_3\phi_T^s(x_2)(\phi_V^s(x_3)(-x_2+x_3+3)-\phi_V^t(x_3)(x_2+x_3-1))\right.\right.\nonumber\\
&&\left.\left.+r_2r_3\phi_T^t(x_2)(\phi_V^s(x_3)(x_2+x_3-1)-\phi_V^t(x_3)(-x_2+x_3-1))\right.\right.\nonumber\\
&&\left.\left.-(1-x_2)\phi_T(x_2)\phi_V(x_3)\right)a_i(t_{van}^1)E_{an}(t_{tan}^1)h_{an}(\sqrt{|\alpha_{an1}^{\prime 2}|},\sqrt{|\beta_{an1}^{\prime 2}|},b_2,b_1)\right.\nonumber\\
&&\left.+\left(-r_2r_3\phi_T^s(x_2)(\phi_V^s(x_3)(-x_2+x_3+1)+\phi_V^t(x_3)(x_2+x_3-1))\right.\right.\nonumber\\
&&\left.\left.+r_2r_3\phi_T^t(x_2)(\phi_V^s(x_3)(x_2+x_3-1)+\phi_V^t(x_3)(-x_2+x_3+1))\right.\right.\nonumber\\
&&\left.\left.+x_3\phi_T(x_2)\phi_V(x_3)\right)a_i(t_{van}^2)E_{an}(t_{tan}^2)h_{an}(\sqrt{|\alpha_{an2}^{\prime
2}|},\sqrt{|\beta_{an2}^{\prime 2}|},b_2,b_1)\right\}\;,
\end{eqnarray}
\item (V-A)(V+A) nonfactorizable annihilation diagrams with the quark entering the tensor meson:
\begin{eqnarray}
F_{tan}^{LR}(a_i)&=&\frac{32}{3}\pi m_B^4 C_F \int_0^1 dx_1 dx_2
dx_3 \int_0^{1/\Lambda_{QCD}}b_1db_1 b_3 db_3
\phi_B(x_1)\nonumber\\
&&\left\{-\left(r_2\phi_V(x_3)(x_2+1)(\phi_T^t(x_2)-\phi_T^s(x_2))+r_3\phi_T(x_2)(x_3-2)(\phi_V^s(x_3)+\phi_V^t(x_3))\right)a_i(t_{van}^1)\right.\nonumber\\
&&\left.\times E_{an}(t_{tan}^1)h_{an}(\sqrt{|\alpha_{an1}^{\prime 2}|},\sqrt{|\beta_{an1}^{\prime 2}|},b_2,b_1)\right.\nonumber\\
&&\left.+\left(r_3x_3\phi_T(x_2)(\phi_V^s(x_3)+\phi_V^t(x_3))-r_2(1-x_2)\phi_V(x_3)(\phi_T^t(x_2)-\phi_T^s(x_2))\right)a_i(t_{van}^2)\right.\nonumber\\
&&\left.\times E_{an}(t_{tan}^2)h_{an}(\sqrt{|\alpha_{an2}^{\prime
2}|},\sqrt{|\beta_{an2}^{\prime 2}|},b_2,b_1)\right\}\;,
\end{eqnarray}
\item (S-P)(S+P) nonfactorizable annihilation diagrams with the quark entering the tensor meson:
\begin{eqnarray}
F_{tan}^{SP}(a_i)&=&\frac{32}{3}\pi m_B^4 C_F \int_0^1 dx_1 dx_2
dx_3 \int_0^{1/\Lambda_{QCD}}b_1db_1 b_3 db_3
\phi_B(x_1)\nonumber\\
&&\left\{\left(r_2r_3\phi_T^s(x_2)(\phi_V^s(x_3)(-x_2+x_3+3)+\phi_V^t(x_3)(x_2+x_3-1))\right.\right.\nonumber\\
&&\left.\left.-r_2r_3\phi_T^t(x_2)(\phi_V^s(x_3)(x_2+x_3-1)+\phi_V^t(x_3)(-x_2+x_3-1))\right.\right.\nonumber\\
&&\left.\left.-x_3\phi_T(x_2)\phi_V(x_3)\right)a_i(t_{van}^1)E_{an}(t_{tan}^1)h_{an}(\sqrt{|\alpha_{an1}^{\prime 2}|},\sqrt{|\beta_{an1}^{\prime 2}|},b_2,b_1)\right.\nonumber\\
&&\left.+\left(-r_2r_3\phi_T^s(x_2)(\phi_V^s(x_3)(-x_2+x_3+1)-\phi_V^t(x_3)(x_2+x_3-1))\right.\right.\nonumber\\
&&\left.\left.+r_2r_3\phi_T^t(x_2)(\phi_V^t(x_3)(-x_2+x_3+1)-\phi_V^s(x_3)(x_2+x_3-1))\right.\right.\nonumber\\
&&\left.\left.-(x_2-1)\phi_T(x_2)\phi_V(x_3)\right)a_i(t_{van}^2)E_{an}(t_{tan}^2)h_{an}(\sqrt{|\alpha_{an2}^{\prime
2}|},\sqrt{|\beta_{an2}^{\prime 2}|},b_2,b_1)\right\}\;.
\end{eqnarray}
\end{itemize}}

Similar to the $B\to VV$ decays, the amplitude of $B\to VT$ can be
decomposed as
\begin{eqnarray}
 {\cal A}(\epsilon_{2},\epsilon_{3})&=&i{\cal A}^N + i(\epsilon^*_{T} \cdot \epsilon^*_{\bullet T}){\cal A}^s
 + (\epsilon_{\mu \nu \alpha \beta}n^{\mu} \bar n^{\nu} \epsilon^{*\alpha}_{T} \epsilon^{*\beta}_{\bullet T}) {\cal A}^p,
\end{eqnarray}
where ${\cal A}^N$ contains the contribution from the longitudinal
polarizations, ${\cal A}^s$ and ${\cal A}^p$ represent the
transversely polarized contributions. With the amplitude functions
obtained in this section, the amplitude for the decay channels can
be expressed. Considering the length of the paper, we will not list
all the expressions of the amplitudes, but give one of the $B$ decay amplitude as
an example:
{\small
 \begin{eqnarray}
 {\cal M}(B^-\to \rho^- a_2^0)=\frac{G_F}{\sqrt{2}}&\bigg\{&
V_{ub}V^*_{ud}[F^{LL}_{vef}(\frac{1}{\sqrt{2}}a_{
1})+F^{LL}_{ven}(\frac{1}{\sqrt{2}}C_{
1})+F^{LL}_{ten}(\frac{1}{\sqrt{2}}C_{
2})+F^{LL}_{vaf}(\frac{1}{\sqrt{2}}a_{ 1}) \nonumber\\ &&
+F^{LL}_{taf}(-\frac{1}{\sqrt{2}}a_{
1})+F^{LL}_{van}(\frac{1}{\sqrt{2}}C_{
1})+F^{LL}_{tan}(-\frac{1}{\sqrt{2}}C_{
1})]-V_{tb}V^*_{td}[F^{LL}_{vef}(\frac{1}{\sqrt{2}}a_{
4}+\frac{1}{\sqrt{2}}a_{10}) \nonumber\\ &&
+F^{LL}_{ven}(\frac{1}{\sqrt{2}}C_{ 3}+\frac{1}{\sqrt{2}}C_{
9})+F^{LL}_{ten}(-\frac{1}{\sqrt{2}}C_{ 3}+\frac{1}{2\sqrt{2}}C_{
9}+\frac{3}{2\sqrt{2}}C_{10}) \nonumber\\ &&
+F^{LR}_{ven}(\frac{1}{\sqrt{2}}C_{ 5}+\frac{1}{\sqrt{2}}C_{
7})+F^{LR}_{ten}(-\frac{1}{\sqrt{2}}C_{ 5}+\frac{1}{2\sqrt{2}}C_{
7})+F^{SP}_{vef}(\frac{1}{\sqrt{2}}a_{ 6}+\frac{1}{\sqrt{2}}a_{ 8})
\nonumber\\ && +F^{SP}_{ten}(\frac{3}{2\sqrt{2}}C_{
8})+F^{LL}_{vaf}(\frac{1}{\sqrt{2}}a_{
4}+\frac{1}{\sqrt{2}}a_{10})+F^{LL}_{taf}(-\frac{1}{\sqrt{2}}a_{
4}-\frac{1}{\sqrt{2}}a_{10}) \nonumber\\ &&
+F^{LL}_{van}(\frac{1}{\sqrt{2}}C_{ 3}+\frac{1}{\sqrt{2}}C_{
9})+F^{LL}_{tan}(-\frac{1}{\sqrt{2}}C_{ 3}-\frac{1}{\sqrt{2}}C_{
9})+F^{LR}_{van}(\frac{1}{\sqrt{2}}C_{ 5}+\frac{1}{\sqrt{2}}C_{ 7})
\nonumber\\ && +F^{LR}_{tan}(-\frac{1}{\sqrt{2}}C_{
5}-\frac{1}{\sqrt{2}}C_{ 7})+F^{SP}_{vaf}(\frac{1}{\sqrt{2}}a_{
6}+\frac{1}{\sqrt{2}}a_{ 8})+F^{SP}_{taf}(-\frac{1}{\sqrt{2}}a_{
6}-\frac{1}{\sqrt{2}}a_{ 8})]
 \bigg\}.
 \end{eqnarray}
}
\section{Numerical results and discussions}
\label{section:Ndata}

With the amplitudes calculated in Sec. \ref{section:Aformula}, the
decay width is given as
\begin{eqnarray}
\Gamma&=&\frac{[(1-(r_2+r_3)^2)(1-(r_2-r_3)^2)]^{1/2}}{16\pi
m_B}\sum_{i}|A_i|^2,
\end{eqnarray}
where $i$ represents all the polarization states, and the branching
ratio is obtained through ${\cal BR}=\Gamma\tau_B$.
The key observables of the decays related in this paper are the CP
averaged branching ratios, polarization fractions, as well as direct
CP asymmetries ($A_{\rm{CP}}^{\rm{dir}}$). Readers are referred to
Ref.~\cite{forcp} for reviews on CP violation. First, we define
four amplitudes as follows:
\begin{eqnarray}
 A_f&=&\langle f|{\cal H}|B\rangle,\;\;\;
 \bar A_f=\langle f|{\cal H}|\bar B\rangle,\nonumber\\
 A_{\bar f}&=&\langle\bar f|{\cal H}|B\rangle,\;\;\;
 \bar A_{\bar f}=\langle\bar f|{\cal H}|\bar B\rangle,
\end{eqnarray}
where $\bar B$ meson has a $b$ quark in it and $\bar f$ is the CP
conjugate state of $f$. The direct CP asymmetry
$A_{\rm{CP}}^{\rm{dir}}$ is defined by
\begin{eqnarray}
 A_{\rm{CP}}^{\rm{dir}}&=&\frac{|\bar A_{\bar f}|^2-|A_f|^2}
                         {|\bar A_{\bar
                         f}|^2+|A_f|^2}\;.\label{eq:Dcp}
\end{eqnarray}
Our results for CP averaged branching ratios and CP asymmetries are
listed in Tables \ref{tab:BmtoVT}, \ref{tab:BdtoVT} and
\ref{tab:BstoVT}. In these tables, we also list the results of the
longitudinal polarization fractions ${\cal R}_L$, which is defined
by
\begin{eqnarray}
 {\cal R}_L&=&\frac{|A_0|^2}{\sum_i |A_i|^2}\;,
\end{eqnarray}
where $A_0$ is the amplitude of the longitudinal polarization. The
first error entries of our results are from the parameters in the
wave functions, the decay constant $f_B$ and the shape parameter
$\omega_b$. The second ones are from $\Lambda_{\rm{QCD}}$, which
varies $20\%$ for error estimates, and from the scale $t$, which are
listed in appendix \ref{appendix:forHD}.
 \begin{table}[h]
\caption{The pQCD results for $B\to VT$ decays which have
experimental data, where the experimental data is from the BaBar
collaboration \cite{Amsler:2008zzb,Aubert:2009sx,Aubert:2008bc}.
Unit $10^{-6}$ for branching ratios, and $\%$ for the ${\cal R}_L$.}
 \label{tab:4channels}
\begin{center}
\begin{tabular}{ccccc}
 \hline\hline
\ \ \          &$\cal BR$  &$$   &${\cal R}_L$  &$$ \\
\ \ \ Decay    &This Work    & Experiments    &This Work   &Experiments\\
\hline
\ \ \ $B^-\to K_2^*(1430)^-\omega$           &$0.81^{+0.62+1.10}_{-0.54-0.62}$   &$21.5\pm 4.3$  &$47.0^{+0.8+0.3}_{-4.2-5.2}$   &$56\pm 11$\\
\ \ \ $\bar B^0\to \bar K_2^*(1430)^0\omega$ &$0.93^{+0.71+1.04}_{-0.51-0.73}$   &$10.1\pm 2.3$  &$55.6^{+3.1+3.0}_{-1.5-3.2}$   &$45\pm12$\\
\ \ \ $B^-\to K_2^*(1430)^-\phi$             &$9.1^{+3.4+2.9}_{-2.6-2.0}$   &$8.4\pm2.1$    &$82.1^{+6.2+8.7}_{-6.6-9.2}$  &$80\pm10$\\
\ \ \ $\bar B^0\to \bar K_2^*(1430)^0\phi$   &$8.7^{+3.1+2.7}_{-2.5-1.9}$   &$7.5\pm1.0$    &$82.0^{+6.5+8.1}_{-6.2-9.7}$   &$90.1^{+5.9}_{-6.9}$\\
\hline \hline
\end{tabular}
\end{center}
\end{table}
 \begin{table}[h]
\caption{The experimental branching ratios of $B\to
K_2^*(1430)(\omega,\phi)$ decays
\cite{Amsler:2008zzb,Aubert:2009sx,Aubert:2008bc} and their
corresponding $B\to VV$ decays \cite{Aubert:2008zza}. The unit is
$10^{-6}$.}
 \label{tab:compare}
\begin{center}
\begin{tabular}{cc|cc}
 \hline\hline
\ \ \  Decays     &$\cal BR$     &Decays     &$\cal BR$ \\
\hline
\ \ \ $B^-\to K_2^*(1430)^-\omega$            &$21.5\pm 4.3$  &$B^-\to K^{*-}\omega$              &$2.4\pm1.0\pm0.2$\\
\ \ \ $\bar B^0\to \bar K_2^*(1430)^0\omega$  &$10.1\pm 2.3$  &$\bar B^0\to \bar K^{*0}\omega$    &$2.0\pm0.5$\\
\ \ \ $B^-\to K_2^*(1430)^-\phi$              &$8.4\pm2.1$    &$B^-\to K^{*-}\phi$                &$10.0\pm2.0$\\
\ \ \ $\bar B^0\to \bar K_2^*(1430)^0\phi$    &$7.5\pm1.0$    &$\bar B^0\to \bar K^{*0}\phi$      &$9.8\pm0.6$\\
 \hline \hline
\end{tabular}
\end{center}
\end{table}

Before we go to the numerical discussions of Tables \ref{tab:BmtoVT}, \ref{tab:BdtoVT} and
\ref{tab:BstoVT}, we note a few comments on the present experimental status.
Only four channels, $B^-\to
 K_2^{*-}(\phi,\omega)$ and $\bar B^0 \to K_2^{*0}(\phi,\omega)$, are
reported by BaBar \cite{Amsler:2008zzb,Aubert:2009sx,Aubert:2008bc},
which are shown in Tables \ref{tab:4channels} and
\ref{tab:compare}. We also collect the corresponding decays in $B\to
VV$ mode \cite{Aubert:2008zza} for comparison. For the helicity
structures of $B\to VT$ decays are very similar to the $B\to VV$
ones, a comparison between $B\to VT$ and $B\to VV$ would be very
enlightening.

Comparing with the experimental data, one can see that the pQCD can
give good predictions for the $B \to \phi (K_2^{*-},\bar K_2^{*0})$
decays. For the $B \to \omega (K_2^{*-},\bar K_2^{*0})$
decays, only the polarization fractions can be accommodated well, and large
deviations exist in the branching ratios. Comparing our predictions with the experimental data, here we would like to make a few comments:
\begin{enumerate}
\item
${\cal BR}(B \to \phi (K_2^{*-},\bar K_2^{*0}))$ is very similar
to ${\cal BR}(B \to \phi (K^{*-},\bar K^{*0}))$, but a little smaller, which
might be understood easily by the effect of a heavier tensor mass on the phase space.
It also indicates that only small effects are brought in the branching ratios when $K^*$ is substituted for $K_2^*$.
\item
However, the experimental data shows totally opposite behavior for the
$B \to \omega (K_2^{*-},\bar K_2^{*0}, K^{*-},\bar K^{*0})$ decays, where
${\cal BR}(B \to \omega (K_2^{*-},\bar K_2^{*0}))$ is much larger than ${\cal BR}(B \to \omega (K^{*-},\bar K^{*0}))$.
In the $B\to VV$ case, ${\cal BR}(B \to \phi (K^{*-},\bar K^{*0}))$ is about five times larger than
${\cal BR}(B \to \omega (K^{*-},\bar K^{*0}))$, while in the $B\to VT$ case ${\cal BR}(B \to \phi (K_2^{*-},\bar K_2^{*0}))$ is even smaller than
${\cal BR}(B \to \omega (K^{*-},\bar K^{*0}))$.
\item
The pQCD predictions for the branching ratios of the $B\to VT$ decays are very
similar to but a little smaller than the experimental data of $B\to VV$. Taking the errors into
consideration, the similar numerical relationship between ${\cal BR}(B \to \phi (K^{*-},\bar K^{*0}))$
and ${\cal BR}(B \to \omega (K^{*-},\bar K^{*0}))$ mentioned above can also be accommodated
in $B\to VT$ decays. As is well known, the $B\to VT$ decay
is very similar to the $B\to VV$ decay mode theoretically, therefore the branching
ratios in these two decay modes are expected to have the similar
behavior.
Based on such prejudice, the present experimental data is a little
difficult to be understood.
\item
However, only BaBar collaboration reported
the results for $B \to \omega (K_2^{*-},\bar K_2^{*0})$ up to now, thus the
experimental data need to be confirmed later. On the theoretical side,
the tensor meson may bring forth new mechanism, which needs
further investigations. In Ref. \cite{Cheng:2010yd} the authors
approached those channels in a different way. They used the
experimental data of those channels to extract the
penguin-annihilation parameters of the QCDF and predicted the other
channels. By adopting the way, the experimental data could also be accommodated.
However, we note more investigations are in need to understand the underlying dynamics
totally.
\end{enumerate}

\begin{table}[b]
\begin{center}
\caption{The branching ratios ($\cal{BR}$ in unit of $10^{-6}$),
polarization fractions (${\cal R}_L$ in unit of $\%$) and direct CP
violation ($A_{\rm{CP}}^{\rm{dir}}$ in unit of $\%$) of $B^-\to VT$
decays.} \label{tab:BmtoVT}
\begin{tabular}{lcccccc}
 \hline\hline
 \ \ \                          &\multicolumn{4}{c}{$\cal{BR}$}         &${\cal{R}}_L$  &$A_{\rm{CP}}^{\rm{dir}}$        \\
 \ \ \ &This Work  & QCDF\cite{Cheng:2010yd}  &ISGW2\cite{Kim:BtoVT}  &CLF\cite{JHMunoz:2009aba}  &This Work   &This Work\\
 \hline
 \ \ \ $B^-\to \omega K_2^{*-}$       &$0.81^{+0.62+1.10}_{-0.54-0.62}$   &$7.5^{+19.7}_{-7.0}$   &$0.112$   &$0.06$    &$47.0^{+0.8+0.3}_{-4.2-5.2}$    &$-4.4_{-0.0-2.2}^{+0.4+3.7}$ \\
 \ \ \ $B^-\to \phi K_2^{*-}$         &$9.1^{+3.4+2.9}_{-2.6-2.0}$   &$7.4^{+25.8}_{-5.2}$   &$2.180$   &$9.24$    &$82.1^{+6.2+8.7}_{-6.6-9.2}$    &$1.5_{-0.1-0.3}^{+0.2+0.1}$ \\
 \ \ \ $B^-\to \rho^- a_2^0$          &$12.8^{+7.1+2.4}_{-5.1-2.4}$       &$8.4^{+4.7}_{-2.9}$   &$7.432$   &$19.34$     &$93.4^{+0.8+1.2}_{-0.9-1.5}$    &$6.5^{+1.4+1.6}_{-1.6-1.5}$ \\
 \ \ \ $B^-\to \rho^- \bar K_2^{*0}$  &$3.9^{+1.2+1.8}_{-0.9-1.0}$        &$18.6^{+50.1}_{-17.2}$   &$$   &$$     &$57.6^{+1.6+1.8}_{-0.8-1.8}$    &$0.43^{+0.50+0.56}_{-0.39-0.07}$ \\
 \ \ \ $B^-\to \rho^0 a_2^-$          &$0.67^{+0.30+0.37}_{-0.20-0.20}$   &$0.82^{+2.30}_{-0.95}$   &$0.007$   &$0.071$    &$50.8^{+5.1+9.5}_{-2.4-8.1}$    &$-6.5^{+0.8+11.2}_{-0.4-7.9}$ \\
 \ \ \ $B^-\to \rho^0 K_2^{*-}$       &$2.3^{+0.6+0.8}_{-0.6-0.5}$        &$10.4^{+18.8}_{-9.2}$   &$0.253$   &$0.74$    &$67.6^{+2.2+1.9}_{-2.9-4.0}$    &$-4.8^{+1.2+1.0}_{-1.8-0.9}$ \\
 \ \ \ $B^-\to \omega a_2^-$          &$0.41^{+0.14+0.07}_{-0.14-0.06}$   &$0.38^{+1.84}_{-0.36}$   &$0.010$   &$0.14$    &$64.5^{+0.5+2.4}_{-2.8-5.1}$    &$5.91^{+2.4+4.2}_{-6.9-7.0}$ \\
 \ \ \ $B^-\to \phi a_2^-$            &$0.01^{+0.01+0.01}_{-0.00-0.00}$   &$0.0003^{+0.013}_{-0.001}$   &$0.004$   &$0.019$    &$67.4^{+0.8+4.3}_{-0.2-0.4}$    &$--$ \\
 \ \ \ $B^-\to K^{*-} a_2^0$          &$3.2^{+1.4+1.1}_{-0.9-0.6}$        &$2.9^{+11.7}_{-2.5}$   &$1.852$   &$2.80$  &$59.4^{+8.1+9.9}_{-7.9-10.7}$    &$-4.1^{+2.9+5.8}_{-3.7-5.2}$ \\
 \ \ \ $B^-\to K^{*-} K_2^{*0}$       &$0.39^{+0.09+0.15}_{-0.10-0.07}$   &$2.1^{+4.2}_{-1.8}$   &$$   &$$    &$68.1^{+0.0+5.3}_{-2.1-2.9}$    &$-3.4^{+0.9+0.9}_{-1.6-5.1}$ \\
 \ \ \ $B^-\to K^{*0} K_2^{*-}$       &$0.19^{+0.08+0.07}_{-0.06-0.05}$   &$0.56^{+1.09}_{-0.38}$   &$0.014$   &$0.59$    &$60.7^{+8.2+6.9}_{-8.8-9.8}$    &$22.1^{+7.5+1.4}_{-7.7-5.2}$ \\
 \ \ \ $B^-\to \bar K^{*0} a_2^-$     &$7.6^{+3.4+2.3}_{-2.7-1.9}$        &$6.1^{+23.8}_{-5.4}$   &$4.495$   &$8.62$    &$61.8^{7.3+8.9}_{-8.1-10.7}$    &$-0.82^{+0.01+0.32}_{-0.25-0.27}$ \\
 \ \ \ $B^-\to \rho^- f_2$            &$15.6^{+8.2+1.8}_{-6.1-2.2}$      &$7.7^{+4.8}_{-2.9}$   &$8.061$   &$$     &$96.9^{+0.0+0.0}_{-0.0-0.0}$    &$7.2^{+0.3+1.2}_{-0.6-1.3}$ \\
 \ \ \ $B^-\to \rho^- f_2^{\prime}$   &$0.11^{+0.04+0.02}_{-0.03-0.02}$   &$0.07^{+0.11}_{-0.04}$   &$0.103$   &$$    &$99.3^{+0.1+0.5}_{-0.0-0.8}$    &$--$ \\
 \ \ \ $B^-\to K^{*-} f_2$            &$7.3^{+2.8+2.4}_{-2.2-1.5}$        &$8.3^{+17.3}_{-6.7}$   &$2.032$   &$$   &$76.3^{+4.1+1.2}_{-3.6-1.5}$    &$-38.6^{+1.7+3.5}_{-0.7-2.7}$ \\
 \ \ \ $B^-\to K^{*-} f_2^{\prime}$   &$1.7^{+0.5+1.0}_{-0.3-0.5}$       &$12.6^{+24.0}_{-11.1}$   &$0.025$   &$$    &$15.1^{+4.2+5.1}_{-3.6-5.6}$    &$-1.6^{+0.6+0.8}_{-1.0-1.2}$ \\
 \hline\hline
\end{tabular}
\end{center}
\end{table}

\begin{table}[h]
\begin{center}
\caption{The branching ratios ($\cal{BR}$ in unit of $10^{-6}$),
polarization fractions (${\cal R}_L$ in unit of $\%$) and direct CP
violation ($A_{\rm{CP}}^{\rm{dir}}$ in unit of $\%$) of $\bar B_d^0
\to VT$ decays.} \label{tab:BdtoVT}
\begin{tabular}{lcccccc}
 \hline\hline
 \ \ \                          &\multicolumn{4}{c}{$\cal{BR}$}         &${\cal{R}}_L$  &$A_{\rm{CP}}^{\rm{dir}}$        \\
 \ \ \ &This Work  & QCDF\cite{Cheng:2010yd}  &ISGW2\cite{Kim:BtoVT}  &CLF\cite{JHMunoz:2009aba}  &This Work   &This Work\\
 \hline
 \ \ \ $\bar B_d^0\to \omega \bar K_2^{*0}$     &$0.93^{+0.71+1.04}_{-0.51-0.73}$   &$8.1^{+21.7}_{-7.6}$   &$0.104$   &$0.053$    &$55.6^{+3.1+3.0}_{-1.5-3.2}$    &$5.4_{-1.7-3.3}^{+1.9+3.5}$ \\
 \ \ \ $\bar B_d^0\to \phi \bar K_2^{*0}$       &$8.7^{+3.1+2.7}_{-2.5-1.9}$   &$7.7^{+26.9}_{-5.5}$   &$2.024$   &$8.51$    &$82.0^{+6.5+8.1}_{-6.2-9.7}$    &$--$ \\
 \ \ \ $\bar B_d^0\to \rho^- a_2^+$             &$26.7^{+13.6+3.5}_{-10.2-3.9}$     &$11.3^{+5.3}_{-4.6}$   &$14.686$   &$36.18$  &$94.8^{+0.3+0.5}_{-0.3-0.6}$    &$1.3^{+0.5+1.0}_{-0.4-0.9}$ \\
 \ \ \ $\bar B_d^0\to \rho^+ a_2^-$             &$0.74^{+0.19+0.16}_{-0.20-0.13}$   &$1.2^{+2.6}_{-1.0}$   &$$   &$$    &$88.7^{+0.8+1.3}_{-1.6-2.6}$    &$7.7^{+3.2+0.6}_{-4.3-4.4}$ \\
 \ \ \ $\bar B_d^0\to \rho^+ K_2^{*-}$          &$3.4^{+1.1+1.6}_{-0.8-0.9}$        &$19.8^{+52.0}_{-18.2}$   &$$   &$$   &$53.9^{+1.0+1.3}_{-0.0-1.4}$    &$-2.2^{+0.7+0.8}_{-0.9-0.8}$ \\
 \ \ \ $\bar B_d^0\to \rho^0 a_2^0$             &$0.68^{+0.30+0.22}_{-0.21-0.12}$   &$0.39^{+1.35}_{-0.20}$   &$0.003$   &$0.03$    &$90.7^{+1.0+1.5}_{-0.0-0.6}$    &$13.6^{+2.1+5.3}_{-0.4-3.1}$ \\
 \ \ \ $\bar B_d^0\to \rho^0 \bar K_2^{*0}$     &$1.7^{+0.5+0.7}_{-0.4-0.4}$       &$9.5^{+33.4}_{-9.5}$   &$0.235$   &$0.68$    &$47.0^{+0.5+2.7}_{-0.8-1.7}$    &$3.3^{+1.6+1.3}_{-1.1-1.9}$ \\
 \ \ \ $\bar B_d^0\to \omega a_2^0$             &$0.37^{+0.11+0.0}_{-0.12-0.03}$   &$0.25^{+1.14}_{-0.19}$   &$0.005$   &$0.07$    &$89.5^{+0.3+1.2}_{-1.6-2.5}$    &$5.7^{+8.9+11.1}_{-7.0-9.9}$ \\
 \ \ \ $\bar B_d^0\to \phi a_2^0$               &$\sim 10^{-3}$                    &$0.001^{+0.006}_{-0.001}$   &$0.002$   &$0.009$    &$45.0^{+1.5+4.8}_{-1.1-3.5}$    &$--$ \\
 \ \ \ $\bar B_d^0\to K^{*-} a_2^+$             &$6.1^{+2.8+2.1}_{-1.7-1.1}$       &$6.1^{+24.3}_{-5.3}$   &$3.477$   &$7.25$    &$59.3^{+8.3+9.1}_{-5.8-9.6}$    &$-17.9^{+1.6+2.6}_{-4.4-4.9}$ \\
 \ \ \ $\bar B_d^0\to K^{*-} K_2^{*+}$          &$3.0^{+0.7+1.4}_{-0.7-0.8}$      &$0.43^{+0.54}_{-0.31}$   &$$   &$$   &$49.8^{+0.3+0.3}_{-1.0-0.8}$    &$5.9^{+3.7+4.3}_{-1.0-0.3}$ \\
 \ \ \ $\bar B_d^0\to K^{*+} K_2^{*-}$          &$3.5^{+0.9+1.7}_{-0.8-0.9}$       &$0.06^{+0.09}_{-0.03}$   &$$   &$$  &$49.4^{+0.3+1.2}_{-0.6-0.4}$    &$-1.2^{+0.4+1.6}_{-0.6-1.2}$ \\
 \ \ \ $\bar B_d^0\to K^{*0}\bar K_2^{*0}$      &$4.5^{+1.3+2.2}_{-1.1-1.2}$       &$0.44^{+0.88}_{-0.30}$   &$0.026$   &$0.55$   &$40.6^{+1.9+2.7}_{-1.5-2.1}$    &$6.1^{+1.4+1.3}_{-0.9-1.9}$ \\
 \ \ \ $\bar B_d^0\to \bar K^{*0} a_2^0$        &$3.5^{+1.7+1.4}_{-1.2-0.8}$       &$3.4^{+12.4}_{-2.8}$   &$2.109$   &$4.03$    &$60.3^{+7.9+9.7}_{-7.7-9.9}$    &$-11.1^{+0.2+2.2}_{-0.1-2.0}$ \\
 \ \ \ $\bar B_d^0\to \bar K^{*0} K_2^{*0}$     &$5.1^{+1.4+2.3}_{-1.2-1.5}$       &$1.1^{+2.9}_{-1.0}$   &$$   &$$    &$55.1^{+0.7+0.0}_{-0.2-1.9}$    &$0.41^{+0.22+0.31}_{-0.19-0.93}$ \\
 \ \ \ $\bar B_d^0\to \rho^0 f_2$               &$0.41^{+0.26+0.15}_{-0.16-0.07}$   &$0.42^{+1.90}_{-0.44}$   &$0.004$   &$0.019$    &$85.3^{+1.7+3.1}_{-1.1-2.0}$    &$-1.9^{+0.0+5.2}_{-2.4-9.4}$ \\
 \ \ \ $\bar B_d^0\to \rho^0 f_2^{\prime}$      &$0.05^{+0.02+0.01}_{-0.01-0.01}$   &$0.03^{+0.06}_{-0.02}$   &$5\times 10^{-5}$   &$$    &$99.3^{+0.0+0.4}_{-0.0-0.8}$    &$--$ \\
 \ \ \ $\bar B_d^0\to \omega f_2$               &$0.56^{+0.19+0.12}_{-0.16-0.11}$   &$0.69^{+0.97}_{-0.36}$   &$0.005$   &$$    &$95.5^{+0.5+0.6}_{-0.5-0.9}$    &$-5.7^{+5.9+7.5}_{-8.3-7.3}$ \\
 \ \ \ $\bar B_d^0\to \omega f_2^{\prime}$      &$0.04^{+0.01+0.01}_{-0.01-0.01}$   &$0.03^{+0.04}_{-0.01}$   &$6\times10^{-5}$   &$$    &$99.2^{+0.0+0.0}_{-0.0-0.0}$    &$--$ \\
 \ \ \ $\bar B_d^0\to \phi f_2$                 &$\sim 10^{-3}$                     &$0.001^{+0.007}_{-0.000}$   &$0.002$   &$$    &$73.0^{+1.12+5.53}_{-0.00-0.00}$    &$--$ \\
 \ \ \ $\bar B_d^0\to \phi f_2^{\prime}$        &$0.06^{+0.03+0.02}_{-0.01-0.05}$   &$0.006^{+0.034}_{-0.005}$   &$2\times10^{-5}$   &$$    &$10.0^{+33.1+24.6}_{-0.0-0.3}$    &$0.67^{+0.0+0.59}_{-4.98-6.03}$ \\
 \ \ \ $\bar B_d^0\to \bar K^{*0} f_2$          &$7.1^{+3.2+2.5}_{-2.1-1.3}$        &$9.1^{+8.8}_{-7.3}$   &$2.314$   &$$    &$73.8^{+5.4+4.3}_{-3.3-1.4}$    &$6.1^{+0.1+1.1}_{-0.4-1.2}$ \\
 \ \ \ $\bar B_d^0\to \bar K^{*0} f_2^{\prime}$ &$1.8^{+0.6+1.1}_{-0.4-0.6}$   &13.5$^{+25.4}_{-11.9}$   &$0.029$   &$$    &$17.4^{+6.7+6.0}_{-1.5-2.5}$    &$--$ \\
 \hline\hline
\end{tabular}
\end{center}
\end{table}

We collect our results for $B_{u,d}\to VT$ decays in the Table
\ref{tab:BmtoVT} and \ref{tab:BdtoVT}, as well as the results of
branching ratios under the QCDF and from the other two models. Most
results of the pQCD and the QCDF agree with each other very well. For those
channels, where the pQCD and the QCDF have obviously different central values,
such as $B^-\to \rho^-\bar K_2^{*0}$ and $\bar B_d^0\to \rho^+\bar
K_2^{*-}$, the penguin-annihilation parameters of the QCDF contribute the differences.
The penguin-annihilation parameters are the key point of the QCDF to
enhance the branching ratios of $B^-\to \omega\bar K_2^{*-}$ and
$\bar B_d^0\to \omega\bar K_2^{*0}$ to accommodate the experimental
data. We presume those parameters are the main factors for the large differences
between the central values of these channels. However, taking
the errors into consideration, the two different theoretical approaches can still agree with each other.
On the other hand, future experimental observation of these channels
may offer an opportunity to test the dynamics of the pQCD and the QCDF.

For the channels dominated by the $W$-emission diagrams, especially
with a vector meson emitted, such as $\bar B_d^0\to a_2^+\rho^-$, the
longitudinal contribution is dominating, and the polarization
fraction ${\cal R}_L$ is around $90\%$. The polarization fractions
of those decays dominated by the penguin diagrams are very profound.
The polarization fractions of some penguin-dominating decays of $B\to
VV$ decay mode like $B\to \phi K^{*0}$ are reported to be around
$50\%$ \cite{Amsler:2008zzb}, which are out of expectation of the SM.
This is so-called the polarization puzzle in
$B$ physics. However, one can find that the polarization fraction of
$B\to\phi K_2^{*0}$ in the $B\to VT$ behaves as the SM expectation,
while the $B\to \omega K_2^{*0}$ gives about $90\%$. In our calculation,
we find that the polarization of $B\to VT$ for these channels are near
to the $B\to VV$ ones. However, after we consider $r^2=(m_T/m_B)^2$
contributions carefully, the polarizations can be accomodated to the
experimental data, although the branching fractions cannot be accomodated.

From Eq. (\ref{eq:Dcp}) one can see that the generation of the
direct CP violation requires that the amplitude $A_f$ consists of at
least two parts with different weak phases. Usually they are the tree
contribution and penguin contributions in the SM. Readers are
referred to Ref. \cite{Amsler:2008zzb} for the related formulas and
reviews. The interference of these parts will bring the direct CP
violation. The magnitude of the direct CP violation is proportional
to the ratio of the penguin and tree contributions. Therefore, the
direct CP violation in the SM is very small, since the penguin
contribution is almost always sub-dominating, which can be seen in previous section.
However, there are a few very special channels in which the
penguin contributions may be comparable to the tree one, as a result,
sizeable direct CP violation appears. Take $B^-\to K^{*-} K_2^{*0}$
as an example. In this channel, the CKM matrix elements for the tree
($V_{ub}V^*_{ud}$) and penguin contributions ($V_{tb}V^*_{td}$) are
at the same order. Although the Wilson coefficients for the tree
contributions are much larger, the tree operator only appears in the
annihilation diagrams, not in the emission ones. Therefore the tree
contributions are suppressed, and the penguin ones become
comparable, which brings to a relatively large direct CP violation
for this channel.
\begin{table}[h]
\begin{center}
\caption{The branching ratios ($\cal{BR}$ in unit of $10^{-6}$),
polarization fractions (${\cal R}_L$ in unit of $\%$) and direct CP
violation ($A_{\rm{CP}}^{\rm{dir}}$ in unit of $\%$) of $\bar B_s^0
\to VT$ decays.} \label{tab:BstoVT}
\begin{tabular}{lccc}
 \hline\hline
 \ \ \                          &$\cal{BR}$             &${\cal{R}}_L$       &$A_{\rm{CP}}^{\rm{dir}}$  \\
 \hline
 \ \ \ $\bar B_s^0\to \rho^- a_2^+$             &$0.35^{+0.08+0.12}_{-0.11-0.08}$    &$75.6^{+0.1+2.9}_{-1.5-3.4}$    &$-11.0^{+3.7+4.9}_{-2.2-2.9}$ \\
 \ \ \ $\bar B_s^0\to \rho^- K_2^{*+}$          &$17.1^{+7.7+1.2}_{-6.0-1.2}$    &$93.5^{+0.2+0.5}_{-0.3-0.6}$    &$4.2^{+0.7+0.8}_{-0.6-1.2}$ \\
 \ \ \ $\bar B_s^0\to \rho^+ a_2^-$             &$0.15^{+0.04+0.08}_{-0.04-0.05}$    &$35.7^{+0.4+6.0}_{-1.2-3.3}$    &$9.2^{+1.1+8.2}_{-1.3-7.6}$ \\
 \ \ \ $\bar B_s^0\to \rho^0 a_2^0$             &$0.03^{+0.01+0.02}_{-0.01-0.01}$    &$99.1^{+0.3+0.6}_{-0.1-1.2}$    &$-18.1^{+3.8+2.5}_{-1.3-5.2}$ \\
 \ \ \ $\bar B_s^0\to \rho^0 K_2^{*0}$          &$0.30^{+0.13+0.14}_{-0.10-0.08}$    &$42.6^{+6.2+8.0}_{-5.9-8.5}$    &$7.0^{+1.4+3.3}_{-2.0-4.7}$ \\
 \ \ \ $\bar B_s^0\to \omega a_2^0$             &$\sim 10^{-3}$    &$59.8^{+12.0+16.7}_{-5.6-13.3}$    &$-18.8^{+1.7+12.7}_{-5.3-3.3}$ \\
 \ \ \ $\bar B_s^0\to \omega K_2^{*0}$          &$0.13^{+0.05+0.07}_{-0.04-0.03}$    &$49.6^{+2.6+6.3}_{-2.1-3.8}$    &$-19.2^{+4.2+6.2}_{-4.8-4.2}$ \\
 \ \ \ $\bar B_s^0\to \phi a_2^0$               &$0.04^{+0.01+0.005}_{-0.01-0.006}$    &$99.2^{+0.1+0.3}_{-0.1-0.6}$    &$-1.1^{+0.1+0.4}_{-1.1-2.2}$ \\
 \ \ \ $\bar B_s^0\to \phi K_2^{*0}$            &$0.36^{+0.10+0.18}_{-0.09-0.10}$    &$62.2^{+3.1+2.3}_{-0.9-0.9}$    &$--$ \\
 \ \ \ $\bar B_s^0\to K^{*-}K_2^{*+}$           &$4.5^{+1.6+1.7}_{-1.1-0.8}$    &$39.9^{+7.8+11.3}_{-4.5-8.3}$    &$-12.5^{+1.2+6.0}_{-0.6-4.3}$ \\
 \ \ \ $\bar B_s^0\to K^{*+}a_2^-$              &$0.66^{+0.18+0.24}_{-0.17-0.15}$    &$77.8^{+1.6+2.6}_{-0.7-3.2}$    &$7.1^{+3.3+3.1}_{-3.0-3.0}$ \\
 \ \ \ $\bar B_s^0\to K^{*+}K_2^{*-}$           &$6.1^{+1.5+2.5}_{-1.6-1.6}$    &$59.9^{+0.3+0.7}_{-1.4-1.9}$    &$-0.9^{+0.7+0.5}_{-0.6-0.4}$ \\
 \ \ \ $\bar B_s^0\to K^{*0}a_2^0$              &$0.88^{+0.25+0.15}_{-0.23-0.13}$    &$90.5^{+0.6+2.3}_{-0.6-3.1}$    &$5.1^{+1.6+2.5}_{-2.6-3.5}$ \\
 \ \ \ $\bar B_s^0\to K^{*0}\bar K_2^{*0}$      &$8.9^{+2.6+3.7}_{-2.2-2.1}$    &$62.9^{+0.4+1.4}_{-1.6-2.9}$    &$--$ \\
 \ \ \ $\bar B_s^0\to \bar K^{*0} K_2^{*0}$     &$6.2^{+1.9+2.2}_{-1.7-1.5}$    &$34.1^{+6.3+11.8}_{-5.2-12.2}$    &$-4.2^{+0.2+0.8}_{-0.3-0.6}$ \\
 \ \ \ $\bar B_s^0\to \rho^0 f_2$               &$\sim 10^{-3}$    &$69.5^{+1.8+8.1}_{-7.1-9.9}$    &$-23.8^{+3.1+5.6}_{-1.6-3.8}$ \\
 \ \ \ $\bar B_s^0\to \rho^0 f_2^{\prime}$      &$0.12^{+0.05+0.01}_{-0.04-0.01}$    &$89.8^{+0.0+0.4}_{-0.0-0.4}$    &$14.2^{+0.8+2.1}_{-0.7-1.9}$ \\
 \ \ \ $\bar B_s^0\to \omega f_2$               &$0.02^{+0.004+0.008}_{-0.003-0.009}$    &$99.2^{+0.1+0.2}_{-0.3-1.0}$    &$-13.9^{+2.6+2.4}_{-6.4-8.1}$ \\
 \ \ \ $\bar B_s^0\to \omega f_2^{\prime}$      &$0.28^{+0.13+0.09}_{-0.10-0.06}$    &$26.4^{+1.7+10.3}_{-0.4-7.4}$    &$-1.3^{+0.5+1.6}_{-0.0-0.5}$ \\
 \ \ \ $\bar B_s^0\to \phi f_2$                 &$2.9^{+1.0+0.7}_{-0.9-0.7}$    &$98.7^{+0.1+0.6}_{-0.0-1.1}$    &$0.84^{+0.07+0.19}_{-0.35-0.41}$ \\
 \ \ \ $\bar B_s^0\to \phi f_2^{\prime}$        &$3.1^{+1.8+0.6}_{-1.4-0.6}$    &$75.3^{+3.0+3.5}_{-3.2-1.7}$    &$--$ \\
 \ \ \ $\bar B_s^0\to K^{*0}f_2$                &$0.51^{+0.17+0.11}_{-0.16-0.13}$    &$92.2^{+1.6+2.4}_{-2.7-5.1}$    &$-11.9^{+4.3+5.9}_{-2.5-2.4}$ \\
 \ \ \ $\bar B_s^0\to K^{*0}f_2^{\prime}$       &$0.39^{+0.13+0.14}_{-0.09-0.08}$    &$59.7^{+3.2+3.3}_{-2.2-3.1}$    &$--$ \\
 \hline\hline
\end{tabular}
\end{center}
\end{table}

For most of the $B\to VT$ decays, the pQCD predicts the branching
ratios at the order of $10^{-6}$, which would be easy for the
experimental observation. We also calculate the branching ratios,
polarization fractions and the direct CP violations of $\bar B_s\to
VT$ decays, which are collected in Table \ref{tab:BstoVT}. Most of
the $\bar B_s$ decays are penguin-dominated, whose branching ratios
are mainly at the order of $10^{-7}$, therefore, whose observation requires more
accumulation of experimental data. However, it would be easy for the
forth-coming future flavor physics experiments. If a vector meson, generated by
the tree operator whose decay constant is nonzero, is emitted in a
$\bar B_s$ decay, then such channels have a large possibility to gain
a relatively large branching ratios with the order of $10^{-6}$.

\section{Summary}
One of the valuable topics in flavor physics is studying the hadrons
in the $B$ meson decays. In recent years, inspired by the
interesting experimental data, more and more studies on the $B$ to
tensor meson decays are carried on. The pQCD approach, which has been being
developed for years and predicts many $B$ meson decays successfully,
is a powerful tool in the study of two body non-leptonic $B$ meson decays.
In this paper, we investigated the $B\to VT$ decays under the frame
of the pQCD. We calculated all the tree level diagrams in the approach
and collected all the necessary expressions in our paper, with which we can study the
$39$ $B\to VT$ and $23$ $B_s\to VT$ decays. The branching ratios,
polarization fractions, and direct CP violations are predicted.

Four channels in $B\to VT$ are reported by the experiments: $B\to
\phi(K_2^{*-}, \bar K_2^{*0})$ and $B\to \omega(K_2^{*-}, \bar
K_2^{*0})$. Comparing with their similar decays in the $B \to VV$
mode, these four channels have very interesting phenomena. On the
experimental side, unlike the polarization puzzle in $B\to VV$
decays, the longitudinal polarization fractions of $B\to
\phi(K_2^{*-}, \bar K_2^{*0})$ decays  are around $90\%$, while
those of the $B\to \omega(K_2^{*-}, \bar K_2^{*0})$ decays are
around $50\%$. The branching ratios of $B\to \omega(K_2^{*-}, \bar
K_2^{*0})$ are much larger than those of $B\to \phi(K_2^{*-}, \bar
K_2^{*0})$. This is quite different from the $B\to VV$ case, where
the branching ratio of $B\to \omega(K^{*-}, \bar K^{*0})$ are about
$5$ times larger than the ones of $B\to \phi(K^{*-}, \bar K^{*0})$.
By considering the $r^2=(m_T/m_B)^2$ corrections, although the
polarization fractions can be accommodated, the branching ratios are
not predicted well. This may need further experimental confirmation
and theoretical investigation.

Most of the branching ratios for $B^-$ and $\bar B^0$ decays are
predicted to be at the order of $10^{-6}$. Most of our results agree
with the the ones of the QCDF. Some channels do not agree so well by the
central values, which may be caused by the different dynamics, since
the QCDF introduce the penguin-annihilation parameters to accommodate
the experimental data and their behavior seems different from the
pQCD approach. However, taking the errors into consideration, they
can still agree. For the decays which contributed by the
$W$-emission diagram, especially when the vector meson is emitted,
the polarization fraction is about $90\%$, which is just as the
expectation of SM. The polarization fractions for the penguin
dominated decays are complicated. Some are around $90\%$ and some
are $50\%$, just like the cases of the four channels observed by the
experiments. Fortunately, the main order of the $B^-$ and $\bar B^0$
decays is $10^{-6}$, which would be easy for the experimental
observation.

In the $\bar B_s^0$ decays, the branching ratios are smaller. Such
tree-dominated decays as $\bar B^0_s \to \rho^- K_2^{*+}$, when a
vector meson is emitted, have the mechanism to gain a relatively
large branching ratio at the order of $10^{-6}$. Most of the others
are at the order of $10^{-7}$, whose observation need more
accumulation of experimental data.

\section{Acknowledgement}
The work was supported by the National Research Foundation of Korea (NRF)
grant funded by Korea government of the Ministry of Education, Science and Technology (MEST)
(Grant No. 2011-0017430) and (Grant No. 2011-0020333). The work of Z.T.Z. was supported by the National Science Foundation of
China under the Grant No.11075168, 11228512, and 11235005.

\begin{appendix}

  \section{Functions for hard kernel, Sudakov factors and scales}
  \label{appendix:forHD}
The parameters in the hard part is given as follows.
\begin{eqnarray}
\beta^2_{ef1} &=& x_1(1-x_2)m_B^2\, ,\, \beta^2_{ef2}=\beta^2_{ef1}\, ,\nonumber\\
\alpha^2_{ef1} &=& (1-x_2)m_B^2\, ,\, \alpha_{ef2}^2=x_1m_B^2\, ,\nonumber\\
\beta^2_{en1} &=& (1-x_2)(x_1-x_3)m_B^2\, ,\,\beta_{en2}^2=(1-x_2)(x_3+x_1-1)m_B^2\, ,\nonumber\\
\alpha^2_{en1} &=& (1-x_2)x_1 m_B^2\, ,\, \alpha_{en2}^2=\alpha^2_{en1}\, ,\nonumber\\
\beta^2_{af1} &=& -x_2(1-x_3)m_B^2\, ,\, \beta^2_{af2}=\beta^2_{af1}\, ,\nonumber\\
\alpha^2_{af1}&=& -x_2 m_B^2\, ,\, \alpha^2_{af2}=(x_3-1)m_B^2\, ,\nonumber\\
\beta^2_{an1} &=& [1-(x_3-x_1)(1-x_2)]m_B^2\, ,\,\beta^2_{an2}=(x_3+x_1-1)x_2 m_B^2\, ,\nonumber\\
\alpha^2_{an1} &=& -x_2(1-x_3)m_B^2\, ,\,\alpha^2_{an2}=\alpha^2_{an1}\, ,\nonumber\\
\beta^{\prime 2}_{i} &=& \beta^2_{i}(x_2\leftrightarrow
x_3)\,,\,\alpha^{\prime 2}_{i}=\alpha^2_{i}(x_2\leftrightarrow
x_3)\, ,
\end{eqnarray}
where $i$ represents any indices.

The functions for the hard parts are given by
\begin{eqnarray}
h_e(\alpha,\beta,b_1,b_2)&=&K_0(\beta
b_2)[\theta(b_2-b_1)I_0(b_1\alpha)K_0(b_2\alpha) +\theta(b_1-b_2)
I_0(b_2\alpha) K_0(b_1\alpha)],\nonumber\\
h_{en}(\alpha,\beta,b_1,b_2)&=&[\theta(b_2-b_1)I_0(b_1\alpha)K_0(b_2\alpha)
+\theta(b_1-b_2) I_0(b_2\alpha) K_0(b_1\alpha)]\times \left\{
\begin{array}{l}
K_0(\beta b_2) \,\,\,\,\,\,\mbox{for }\beta^2>0\\
\frac{i\pi}{2}H^{(1)}_0(\beta b_2)\,\,\,\mbox{for }\beta^2<0
\end{array}\right\},\nonumber\\
h_a(\alpha,\beta,b_1,b_2)&=&\left(\frac{i\pi}{2}\right)^2
H_0^{(1)}(\beta
b_2)[\theta(b_2-b_1)J_0(b_1\alpha)H^{(1)}_0(b_2\alpha)
+\theta(b_1-b_2) J_0(b_2\alpha) H^{(1)}_0(b_1\alpha)],\nonumber\\
h_{an}(\alpha,\beta,b_1,b_2)&=&\frac{i\pi}{2}[\theta(b_2-b_1)J_0(b_1\alpha)H^{(1)}_0(b_2\alpha)\nonumber\\
&&+\theta(b_1-b_2) J_0(b_2\alpha) H^{(1)}_0(b_1\alpha)]\times\left\{
\begin{array}{l}
K_0(\beta b_2) \,\,\,\,\,\,\mbox{for }\beta^2>0\\
\frac{i\pi}{2}H^{(1)}_0(\beta b_2)\,\,\,\mbox{for }\beta^2<0
\end{array}\right\},
\end{eqnarray}
where the $K_0$, $I_0$, $J_0$ and $H_0^{(1)}$ are all Bessel
functions, and $H_0^{(1)}(z)=J_0(z)+i Y_0(z)$.

The scales are defined as
\begin{eqnarray}
t_{v..}^l=\max\left[c\sqrt{|\alpha_{..l}^2|},c\sqrt{|\beta_{..l}^2|},1/b_k,1/b_l\right],\;
t_{t..}^l=\max\left[c\sqrt{|\alpha_{..l}^{\prime
2}|},c\sqrt{|\beta_{..l}^{\prime 2}|},1/b_k,1/b_l\right],
\end{eqnarray}
where $l=1,2$, the "$..$" represents $ef$, $en$, $af$ or $an$, and
$b_{k,l}$ represent the two corresponding $b$ coordinates in the
measurement of the integration. The parameter $c=1$, and in our
error estimation, we choose $c=0.75$ and $1.25$ for a rough
estimation.

The expressions for the Sudakov factors and coupling constants are
given as
 \begin{eqnarray}
 E_e(t)&=&\alpha_s(t)\exp[-S_B(t)-S_V(t)]\;,\nonumber \\
 E_a(t)&=&\alpha_s(t)\exp[-S_T(t)-S_V(t)]\;,\nonumber\\
 E_{en}(t)&=&\left\{
 \begin{array}{l}
 \alpha_s(t)\exp[-S_B(t)-S_T(t)-S_V(t)|_{b_2=b_1}],\,\,\mbox{if vector meson emits}
 \\
 \alpha_s(t)\exp[-S_B(t)-S_T(t)-S_V(t)|_{b_3=b_1}],\,\,\mbox{if tensor meson emits}
 \end{array}
 \right\}
 \;,\nonumber \\
 E_{an}(t)&=&\alpha_s(t)\exp[-S_B(t)-S_T(t)-S_V(t)|_{b_3=b_2}]\;,
 \end{eqnarray}
where
 \begin{eqnarray}
 S_B(t)&=&s(x_1\frac{m_{B}}{\sqrt{2}},b_1)+\frac{5}{3}\int^t_{1/b_1}\frac{d\bar \mu}{\bar
 \mu}\gamma_q(\alpha_s(\bar \mu)),\nonumber\\
 S_T(t)&=&s(x_2\frac{m_{B}}{\sqrt{2}},b_2)+s((1-x_2)\frac{m_{B}}{\sqrt{2}},b_2)+2\int^t_{1/b_2}\frac{d\bar \mu}{\bar
 \mu}\gamma_q(\alpha_s(\bar \mu)),\nonumber\\
 S_V(t)&=&s(x_3\frac{m_{B}}{\sqrt{2}},b_3)+s((1-x_3)\frac{m_{B}}{\sqrt{2}},b_3)+2\int^t_{1/b_3}\frac{d\bar \mu}{\bar
 \mu}\gamma_q(\alpha_s(\bar \mu)),
 \end{eqnarray}
 with the quark anomalous dimension $\gamma_q=-\alpha_s/\pi$. The
explicit form for the  function $s(Q,b)$ is:
\begin{eqnarray}
s(Q,b)&=&~~\frac{A^{(1)}}{2\beta_{1}}\hat{q}\ln\left(\frac{\hat{q}}
{\hat{b}}\right)-
\frac{A^{(1)}}{2\beta_{1}}\left(\hat{q}-\hat{b}\right)+
\frac{A^{(2)}}{4\beta_{1}^{2}}\left(\frac{\hat{q}}{\hat{b}}-1\right)
-\left[\frac{A^{(2)}}{4\beta_{1}^{2}}-\frac{A^{(1)}}{4\beta_{1}}
\ln\left(\frac{e^{2\gamma_E-1}}{2}\right)\right]
\ln\left(\frac{\hat{q}}{\hat{b}}\right)
\nonumber \\
&&+\frac{A^{(1)}\beta_{2}}{4\beta_{1}^{3}}\hat{q}\left[
\frac{\ln(2\hat{q})+1}{\hat{q}}-\frac{\ln(2\hat{b})+1}{\hat{b}}\right]
+\frac{A^{(1)}\beta_{2}}{8\beta_{1}^{3}}\left[
\ln^{2}(2\hat{q})-\ln^{2}(2\hat{b})\right],
\end{eqnarray} where the variables are defined by
\begin{eqnarray}
\hat q\equiv \mbox{ln}[Q/(\sqrt 2\Lambda)],~~~ \hat b\equiv
\mbox{ln}[1/(b\Lambda)], \end{eqnarray} and the coefficients
$A^{(i)}$ and $\beta_i$ are \begin{eqnarray}
\beta_1=\frac{33-2n_f}{12},~~\beta_2=\frac{153-19n_f}{24},\nonumber\\
A^{(1)}=\frac{4}{3},~~A^{(2)}=\frac{67}{9}
-\frac{\pi^2}{3}-\frac{10}{27}n_f+\frac{8}{3}\beta_1\mbox{ln}(\frac{1}{2}e^{\gamma_E}),
\end{eqnarray}
$n_f$ is the number of the quark flavors and $\gamma_E$ is the Euler
constant. We will use the one-loop running coupling constant, i.e.
we pick up the four terms in the first line of the expression for
the function $s(Q,b)$.

\end{appendix}

\newpage


\end{document}